\documentclass[12pt]{iopart}

\usepackage{iopams}
\usepackage[dvipdfm]{graphicx}

\begin{document}

\title[Spin waves in dipole rings]
{Spin waves in rings of classical magnetic dipoles}
\author{Heinz-J\"urgen Schmidt$^1$, Christian Schr\"oder$^2$, and Marshall Luban$^3$
}
\address{$^1$Department of Physics, University of Osnabr\"uck,
 D - 49069 Osnabr\"uck, Germany\\
$^2$Bielefeld Institute for Applied Materials Research, Bielefeld University
of Applied Sciences, D - 33692 Bielefeld,
Germany \\
$^3$Department of Physics and Astronomy, Iowa State University,
Ames, IA 50011, USA }

\ead{hschmidt@uos.de}
\vspace{10pt}

\begin{abstract}
We theoretically and numerically investigate spin waves that occur in systems of
classical magnetic dipoles that are arranged at the vertices of a regular
polygon and interact solely via their magnetic fields. There are certain limiting cases that can be analyzed in detail. One case is that of
spin waves as infinitesimal excitations from the system's ground state, where the dispersion relation can be determined analytically.
The frequencies of these infinitesimal spin waves are compared with the peaks of the Fourier transform of the thermal expectation value
of the autocorrelation function calculated by Monte Carlo simulations.
In the special case of vanishing wave number an exact solution of the equations of motion is possible describing synchronized oscillations with finite amplitudes.
Finally, the limiting case of a dipole chain with $N\longrightarrow \infty$ is investigated and completely solved.
\end{abstract}

%
\vspace{2pc}

%
%
%
%

\maketitle

\section{Introduction}\label{sec:I}

Systems in which magnetic nanostructures solely interact via
electromagnetic forces have recently drawn much attention
experimentally as well as theoretically \cite{EW13} -- \cite{MCT16}.
Whereas in traditional magnetic systems electromagnetic forces
usually just add to a complex exchange interaction scenario, they
play a major role in arrays of interacting magnetic nanoparticles
and lithographically produced nanostructures, like nanodots and nanopillars.
In such systems geometrical frustration and disorder lead to interesting and exotic
low temperature effects, e.~g.~artificial spin ice  \cite{W06,C08},
and superspin glass behavior \cite{H11}.
Moreover, the dynamic behavior of interacting magnetic nanostructures is a
subject of considerable current investigation.
These systems are promising candidates for future applications beyond
magnetic data-storage, e.~g.~, as low-power logical devices
\cite{I06,E14}. Theoretically, these systems can often be described as
interacting point dipoles. This is justified if the considered
nanostructures form single domain magnets and are spatially well
separated from each other so that exchange interactions do not play
an important role.\\
A different realization of systems of interacting dipoles is given by $N@C_{60}$,
nitrogen atoms encapsulated in fullerene cages \cite{W00}. Such fullerene array structures
have been proposed as an alternative concept for a scalable spin quantum computer \cite{H02}.

The basic building block of all extended interacting dipole systems represents a pair of interacting magnetic dipoles with fixed position. This system has been shown \cite{SSHL15} to be completely integrable due to the existence of two conserved quantities, namely the energy $H$ and the component ${\mathbf S}\cdot{\mathbf e}$ of the total magnetic moment ${\mathbf S}$ into the direction ${\mathbf e}$ joining the two dipoles. The two ground states of the dipole pair
are those where both moment vectors are parallel to ${\mathbf e}$ or to $-{\mathbf e}$ and thus ferromagnetic in character. Larger systems of magnetic dipoles will no longer be integrable and it will be, in general, difficult to determine their ground states because of geometric frustration.
However, for special geometries the ground states may be found, see, e.~g.~\cite{SSR15}.

\begin{figure}
\begin{center}
\includegraphics[clip=on,width=120mm,angle=0]{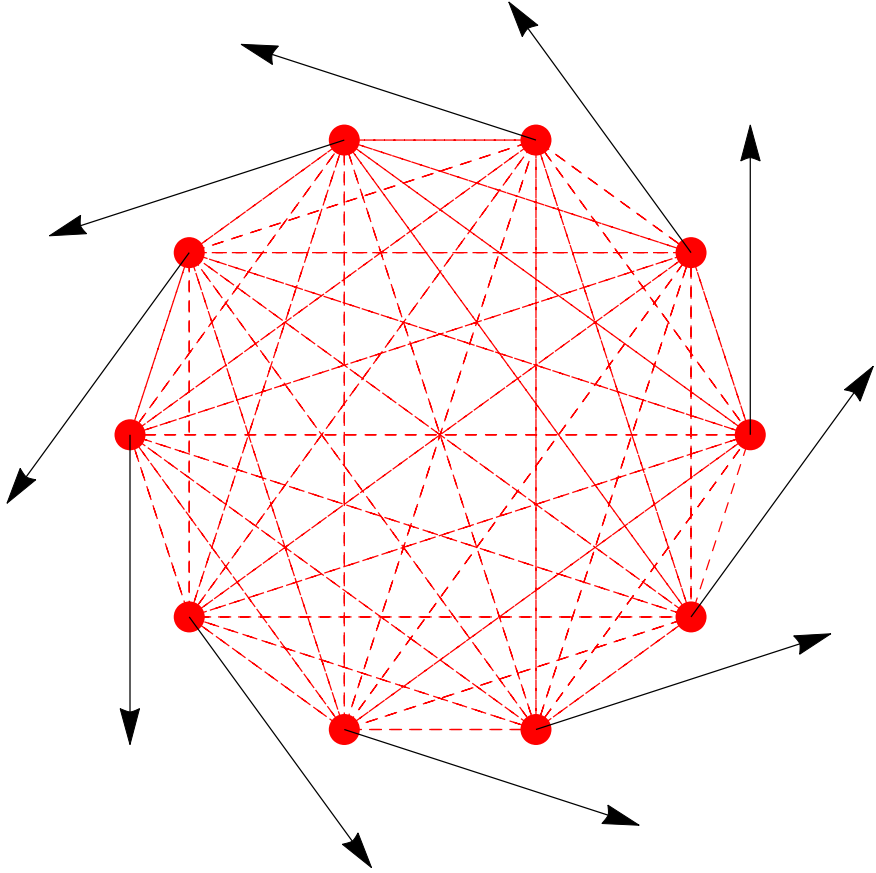}
\end{center}
\caption{Illustration of one of the two ground states $\pm \breve{\mathbf s}$ of the $N=10$ dipole ring.
\label{figGS}
}
\end{figure}

In this paper we consider {\it discrete dipole rings}, i.~e.~, systems where the dipoles are located
at the vertices of a regular polygon, see section \ref{sec:R}, and, as a limit case, the infinite dipole chain.
For the latter no frustration occurs and the two ferromagnetic ground states are obvious.
For dipole rings numerical evidence shows that the two ground states $\pm\breve{\mathbf s}$ are those with all moment vectors
tangent to the circle circumscribing the polygon, see figure \ref{figGS}. For a rigorous proof of the ground state property
of $\pm\breve{\mathbf s}$ see \cite{S16} and some closely related remarks in \ref{sec:A}.
It is then a straightforward task to linearize the equation of motion (eom)
just above the ground state energy and to solve it, see section \ref{sec:L}.
Due to the $C_N$ symmetry of the dipole ring these solutions can be viewed as {\it spin waves},
where the notation has been chosen in analogy to similar solutions for classical spin lattices \cite{M81}.
The frequencies of the linearized spin waves can also be numerically determined, see section \ref{sec:S}:
We calculate the autocorrelation function $\langle\mbox{ac}(t)\rangle$ in the thermal average for low temperatures and consider
the maxima of its Fourier transform.
The position of the dominant peaks then coincides with the theoretically determined frequencies
within a precision of $0.25$ percent for the cases $N=3,\ldots 7$ that we have investigated.

\begin{figure}
\centering
	\includegraphics[width=0.4\linewidth]{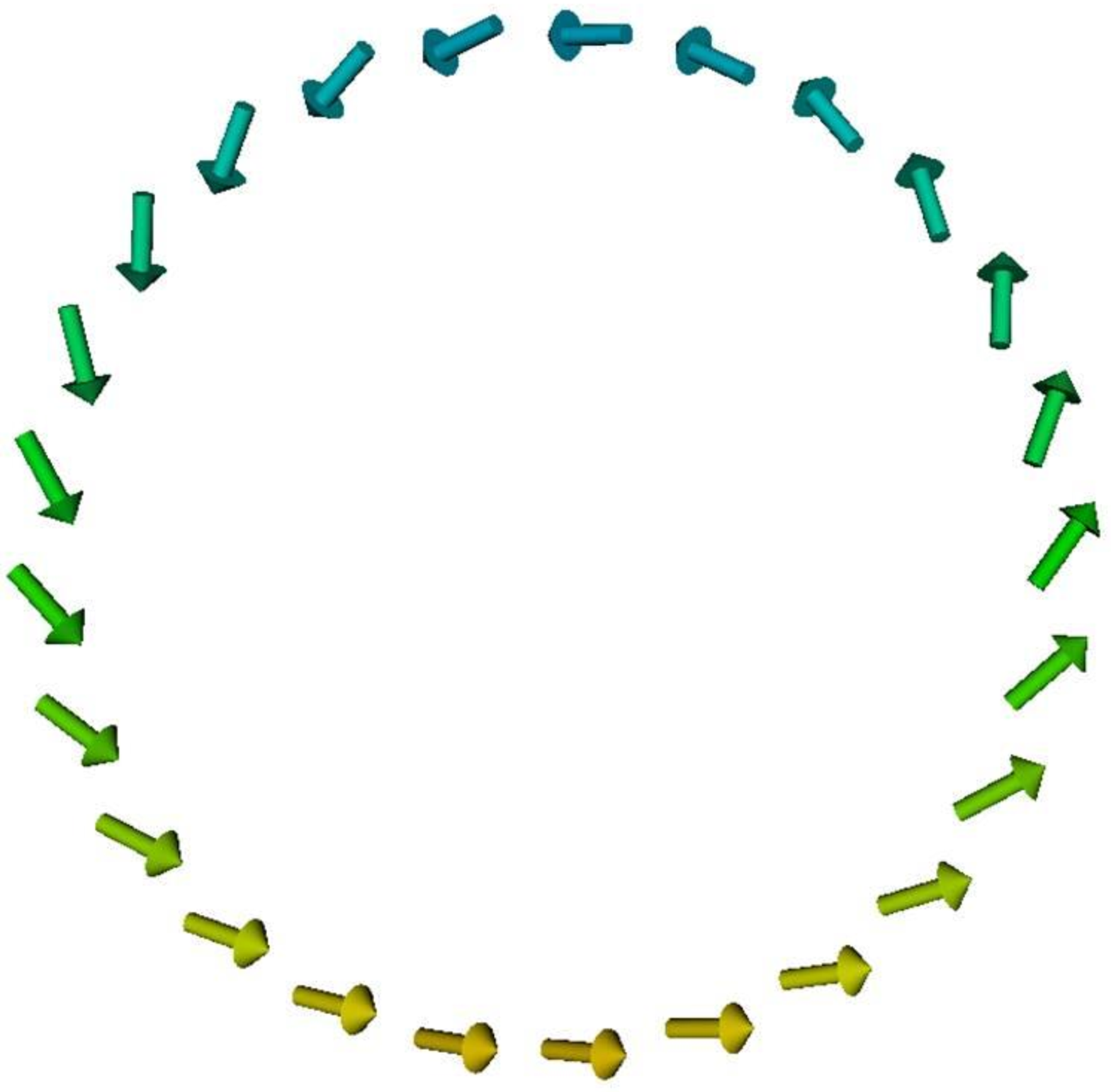}
	\includegraphics[width=0.4\linewidth]{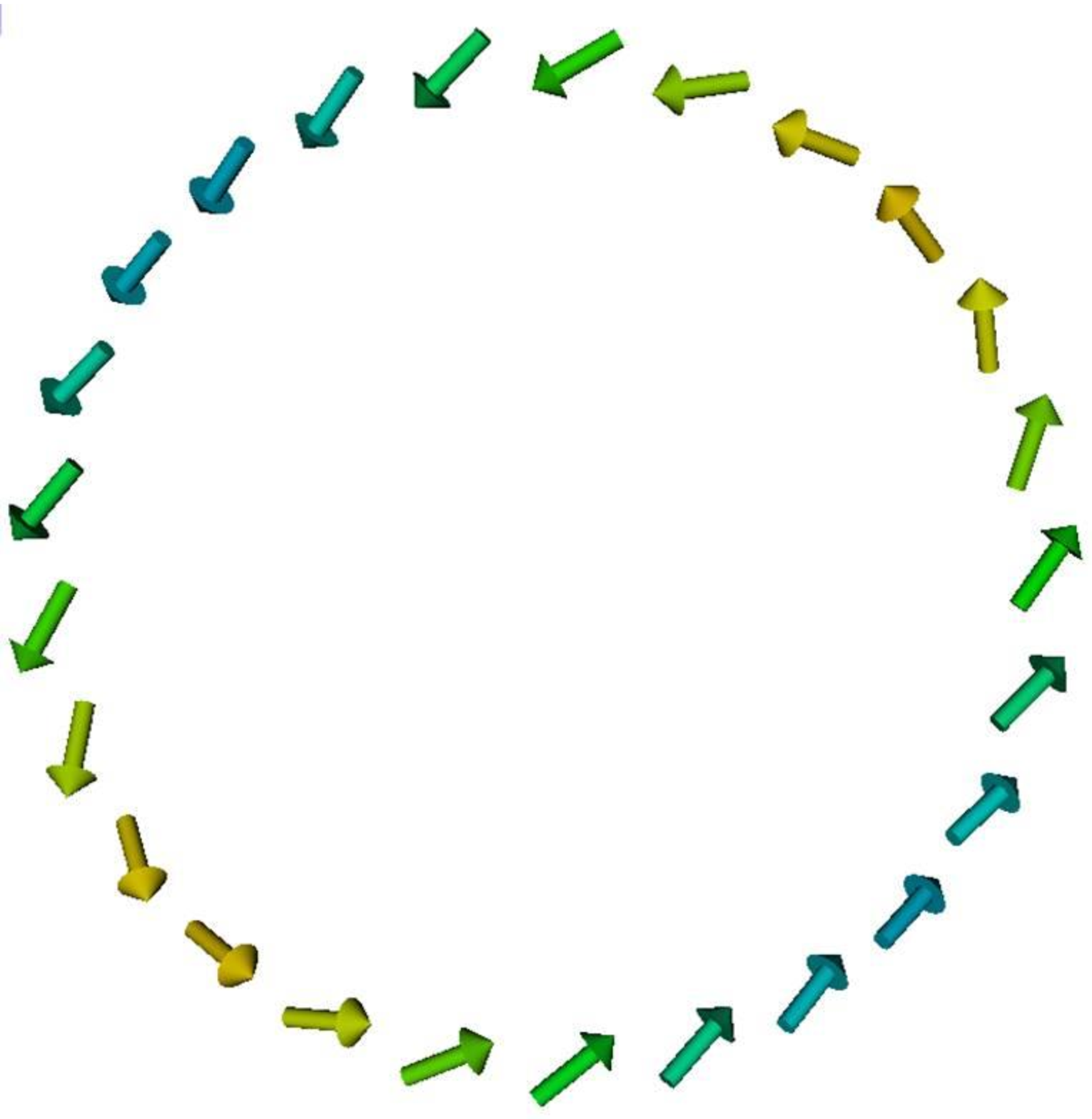}
	\includegraphics[width=0.4\linewidth]{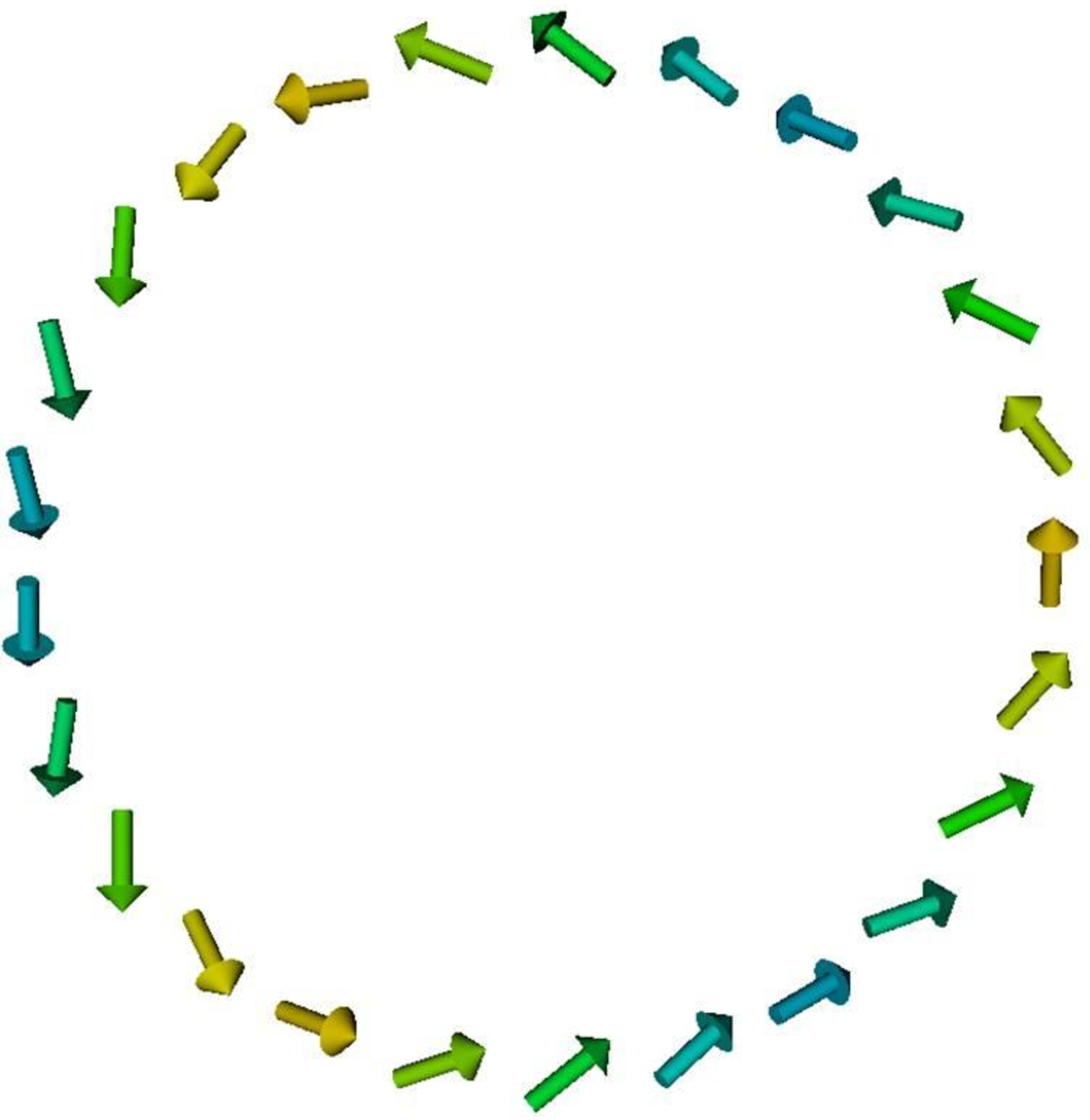}
	\includegraphics[width=0.4\linewidth]{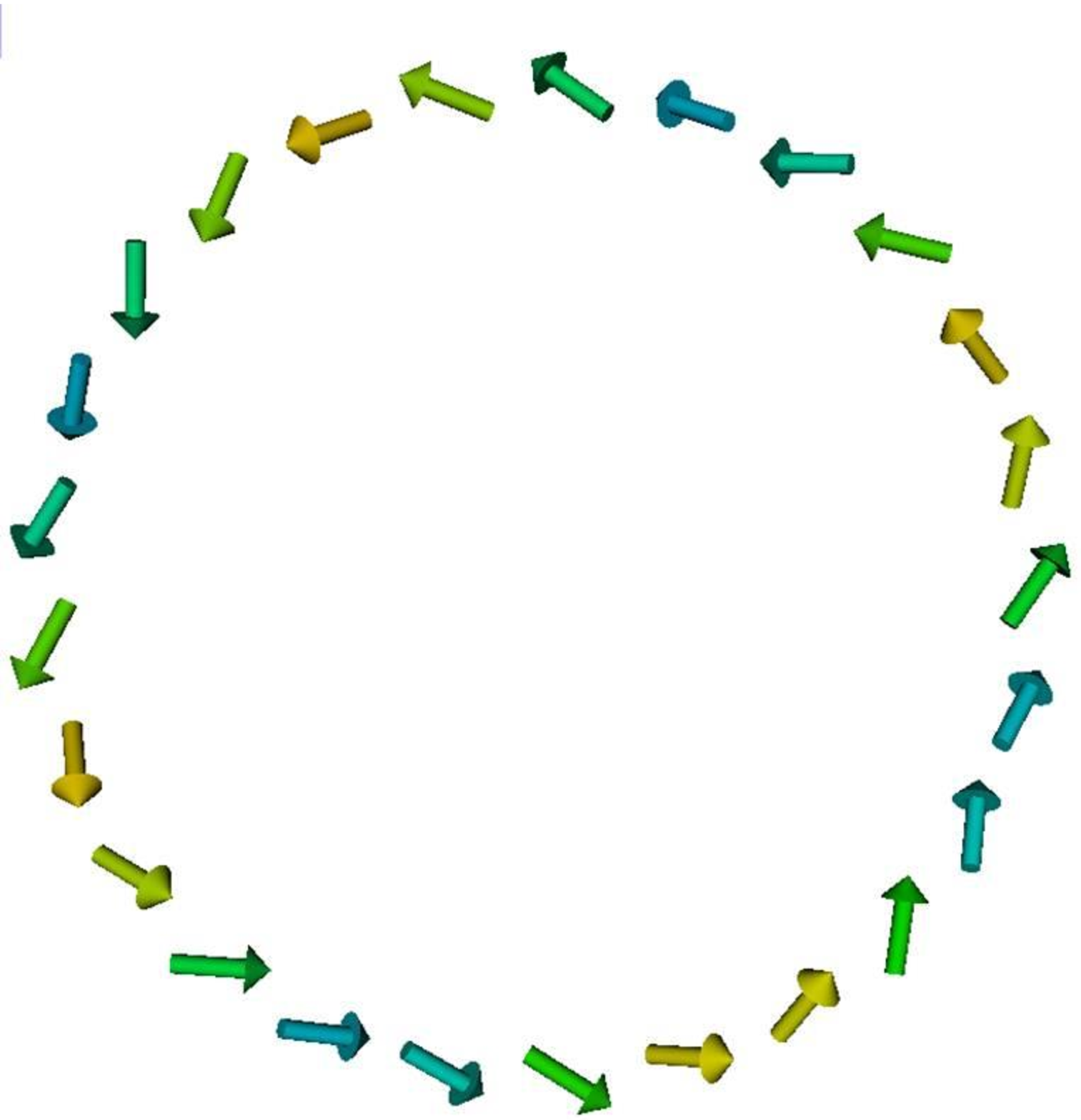}
\caption{The first four spin wave modes $k_\mu,\;\mu=1,2,3,4$ for a ring of $25$ coupled dipoles. The colour coding helps to visualize the wave pattern.
\label{fig_25spinring}
}
\end{figure}

Spin waves are interesting phenomena in their own right as well as tools for an approximate
description of quantum systems \cite{A97}. The simplest models of spin waves can be obtained for a spin ring
with classical nearest neighbor Heisenberg interaction \cite{M81}. It shows not only infinitesimal spin waves in the sense
sketched above but also finite amplitude spin waves with a dispersion relation \cite{SSL11}
\begin{equation}\label{I1}
\omega(q)=2(1-\cos q)\,z
\;.
\end{equation}
The dipole ring is more complicated in two respects: The interaction is (1) anisotropic
in dependence of the direction ${\mathbf e}_{\mu\nu}$ joining any two dipoles and (2)
of long range. Thus the question arises whether also for dipole rings finite amplitude spin waves exist.

The solutions to the linearized eom described above
can be re-translated into approximate finite amplitude solutions via some inverse projection.
 Alternatively, we have constructed the corresponding finite amplitude initial conditions
and numerical solved the eom with these conditions. This yields numerically exact solutions
that are, however, not spin waves in a strict sense. Of course, the quality of these
``approximate spin waves" depends on the size of the initial amplitudes: The smaller
the amplitudes, the better the approximation of spin waves. 
Moreover, the quality increases with N. We have found numerical spin waves from $N=3,\ldots,50$. 
For example,  Fig.~\ref{fig_25spinring} shows a numerical solution for $N=25$ with large amplitudes that looks perfectly
like a spin wave.\\
Given this situation the above question assumes the following form: Are finite amplitude solutions always
only spin waves in some approximate sense, or do there exist exact finite amplitude spin waves?
We cannot definitely answer this question but we have strong evidence that exact spin waves exist.
There are at least three arguments in favor of this.
\begin{enumerate}
\item In the case of $N=3$ we have systematically improved the initial conditions obtained from
solutions of the
linearized eom and obtained numerical solutions that are rather close to exact spin waves, see section \ref{sec:F}
\item In the special case of vanishing wave number $k=0$ it is possible to obtain analytical solutions that describe
``synchronized oscillations", see section \ref{sec:SO}. It would seem to be strange if such exact spin waves exist for $k=0$
but not for larger $k$.
\item In the limit of $N\longrightarrow\infty$, i.~e.~, for the dipole chain, exact spin waves exist with arbitrary amplitudes, see section
\ref{sec:DC}. Again, it would seem to be strange if such exact spin waves exist for $N\longrightarrow\infty$
but not for finite $N$.
\end{enumerate}
These arguments are further discussed in section \ref{sec:S} containing the summary and the outlook for future investigations.
Interestingly, in the case of the infinite dipole chain there exist spin waves with negative group velocity, see section
\ref{sec:DC}.
Systems with negative group velocity have recently found much attention in optics, see, e.~g.~\cite{AT16}, as well as in the investigation of meta-materials, see, e.~g.~\cite{LX16}, \cite{MG16}.

\section{Rings of interacting magnetic dipoles}
\label{sec:R}
We consider systems of $N$ classical point-like dipoles. The normalized dipole
moments are described by unit vectors ${\mathbf s}_\nu,\;\nu=0,\ldots,N-1$.
Each dipole moment performs a precession about the instantaneous magnetic field vector
that results as a sum over all magnetic fields produced by the other dipoles.
The $N$ dipoles are fixed at the positions of the vertices of a regular $N-$ polygon
\begin{equation}\label{R1}
{\mathbf r}_\nu =
\left(
\begin{array}{c}
 \cos \frac{2\pi\nu}{N} \\
 \sin \frac{2\pi\nu}{N} \\
 0\\
\end{array}
\right),\quad \nu=0,\ldots,N-1
\;.
\end{equation}
For the sake of simplicity, the length of the vectors ${\mathbf r}_\nu$ is chosen as $1$, but it can be scaled arbitrarily.
Then the dimensionless equation of motion (eom)
assumes the form
\begin{equation}\label{R2}
\frac{d}{dt}{\mathbf s}_\nu = {\mathbf s}_\nu \times \sum_{\scriptsize\begin{array}{c}\mu=0,\ldots N-1\\ \mu\neq \nu\end{array}}
\frac{1}{|{\mathbf r}_\mu-{\mathbf r}_\nu|^3}
\left( {\mathbf s}_\mu-3\,{\mathbf s}_\mu\cdot {\mathbf e}_{\mu \nu}\;{\mathbf e}_{\mu \nu}
\right)
\;,
\end{equation}
where
\begin{equation}\label{R3}
{\mathbf e}_{\mu \nu}\equiv \frac{{\mathbf r}_\mu-{\mathbf r}_\nu}{|{\mathbf r}_\mu-{\mathbf r}_\nu|} \mbox{ for } \mu\neq\nu
\;.
\end{equation}
For a detailed derivation and analysis of this eom that apply to the special case of $N=2$ see \cite{SSHL15}.
The time evolution for the lapse of time $t$ will be denoted by the operator $T_t: {\mathcal P}\longrightarrow {\mathcal P}$,
where ${\mathcal P}$ denotes the phase space of the dipole ring, see section \ref{sec:L} for more details.
The dimensionless energy of the dipole system is
\begin{equation}\label{R4}
H=\sum_{\scriptsize\begin{array}{c}\mu,\nu=0,\ldots N-1\\ \mu\neq \nu\end{array}}
\frac{1}{|{\mathbf r}_\mu-{\mathbf r}_\nu|^3}\;
{\mathbf s}_\nu\cdot
\left( {\mathbf s}_\mu-3\,{\mathbf s}_\mu\cdot {\mathbf e}_{\mu \nu}\;{\mathbf e}_{\mu \nu}
\right)
\;,
\end{equation}
see \cite{G99} (6.35), \cite{SSHL15}. It will be conserved under time evolution.
By definition, the {\it ground states} of the system are spin configurations that assume the global minimum of
the energy (\ref{R4}). Numerical studies \cite{VD14} suggest that there are exactly two ground states, namely
\begin{equation}\label{R5}
\breve{\mathbf s}_\nu=\left(
\begin{array}{c}
 -\sin \frac{2\pi\nu}{N} \\
 \cos \frac{2\pi\nu}{N} \\
 0
\end{array}
\right),\quad \nu=0,\ldots,N-1
\;,
\end{equation}
and $-\breve{\mathbf s}_\nu,\;\nu=0,\ldots,N-1$, see Fig.~\ref{figGS} for an illustration.
The rigorous proof of this suggestion can be found in \cite{S16}; we will confine ourselves to some remarks in \ref{sec:A}
that support the ground state property of (\ref{R5}).

We define the ``shift operator" $S: {\mathcal P}\longrightarrow {\mathcal P}$ as a cyclic permutation of the moment sites
accompanied by a rotation $R_3(\alpha)$ about the $3$-axis:
\begin{equation}\label{R6}
(S\,{\mathbf s})_\nu \equiv R_3\left(\frac{2\pi}{N}\right){\mathbf s}_{\nu+1}
\;,\quad \nu=0,\ldots,N-1
\;,
\end{equation}
where the addition $\nu +1$ is understood modulo $N$.
This shift is a symmetry operation of the dipole ring in the sense that it commutes with the
time evolution:
\begin{equation}\label{R7}
S\; T_t= T_t\; S\quad \mbox{ for all }t
\;.
\end{equation}

\section{Linearized spin waves}
\label{sec:L}
For excitations of the dipole ring slightly above the ground state energy it is
sensible to linearize the eom at the ground state $\breve{\mathbf s}$ given by (\ref{R5}).
Some mathematical considerations are in order. The eom (\ref{R2}) can be viewed as a
Hamiltonian eom on a $2N$-dimensional phase space that is the $N$-fold product of unit
spheres, ${\mathcal P}={\mathcal S}^2\times\ldots\times {\mathcal S}^2$, see \cite{SSHL15}.
Linearization of this eom takes place in the $2N$-dimensional tangent space $T_{\breve{\mathbf s}}{\mathcal P}$
at the point $\breve{\mathbf s}\in{\mathcal P}$. This tangent space can be visualized as
the $N$-fold product of planes $T_\mu$ tangent to the unit sphere ${\mathcal S}^2$ at the points $\breve{\mathbf s}_\mu$.
We introduce the basis $({\mathbf e}_3,{\mathbf r}_\mu)$ in the tangent plane $T_\mu$, where
${\mathbf e}_3=(0,0,1)^\top$
and consider the coordinates $(\zeta_\mu,\xi_\mu)$ of $T_\mu$ w.~r.~t.~this basis.
This yields coordinates $(\zeta,\xi)\equiv(\zeta_0\ldots,\zeta_{N-1},\xi_0,\ldots,\xi_{N-1})$
of $T_{\breve{\mathbf s}}{\mathcal P}$.
A small neighborhood of ${\mathbf 0}\in T_{\breve{\mathbf s}}{\mathcal P}$
can be identified with a small neighborhood of $\breve{\mathbf s}\in{\mathcal P}$
by means of the projection
\begin{equation}\label{L1}
{\mathbf s}_\mu \mapsto (\zeta_\mu,\xi_\mu)=({\mathbf e_3}\cdot{\mathbf s}_\mu,{\mathbf r}_\mu\cdot{\mathbf s}_\mu),\;\mu=0,\ldots,N-1
\;,
\end{equation}
that is locally $1:1$. In this sense, the eom on ${\mathcal P}$ can be
translated into an eom on $T_{\breve{\mathbf s}}{\mathcal P}$ and, further, linearized
w.~r.~t.~the coordinates $(\zeta,\xi)$ by means of a Taylor expansion. Note that the constant
term of the Taylor expansion vanishes since $\breve{\mathbf s}$ is a stationary point of the eom.
Then the linearized eom assumes the form
\begin{equation}\label{L2}
\frac{d}{dt}\left(
\begin{array}{c}
 \zeta \\
 \xi \\
\end{array}
\right)={\mathbf M}
\left(
\begin{array}{c}
 \zeta \\
 \xi \\
\end{array}
\right)
=
\left(
\begin{array}{cc}
 0&M^{(1)} \\
 M^{(2)}&0 \\
\end{array}
\right)
\left(
\begin{array}{c}
 \zeta \\
 \xi \\
\end{array}
\right)
\;.
\end{equation}
The two $N\times N$ sub-matrices $M^{(i)},\; i=1,2$ are symmetric circulants and hence commute with each other.
A {\it circulant} is a square matrix that commutes with the cyclic shift matrix, see \cite{A01}, part III.
For example, in the case $N=4$ the matrix $M^{(1)}$ has the following form
\begin{equation}\label{L3}
\left(
\begin{array}{cccc}
 \frac{1}{8} \left(1+6 \sqrt{2}\right) & \frac{1}{2 \sqrt{2}} & \frac{1}{8} & \frac{1}{2 \sqrt{2}} \\
 \frac{1}{2 \sqrt{2}} & \frac{1}{8} \left(1+6 \sqrt{2}\right) & \frac{1}{2 \sqrt{2}} & \frac{1}{8} \\
 \frac{1}{8} & \frac{1}{2 \sqrt{2}} & \frac{1}{8} \left(1+6 \sqrt{2}\right) & \frac{1}{2 \sqrt{2}} \\
 \frac{1}{2 \sqrt{2}} & \frac{1}{8} & \frac{1}{2 \sqrt{2}} & \frac{1}{8} \left(1+6 \sqrt{2}\right) \\
\end{array}
\right)
\;.
\end{equation}
One notes that all secondary diagonals have equal entries even if the diagonals are ``periodically extended".
The eigenvectors of a circulant form the discrete Fourier basis and its list of eigenvalues is the discrete Fourier transform
of the first matrix row (times the factor $\sqrt{N}$), see \cite{A01}. In our case, the circulant property
of $M^{(i)},\; i=1,2$ is clearly the consequence of the $C_N$- symmetry of the dipole ring. The symmetry
of $M^{(i)},\; i=1,2$ can be traced back to the property that ${\mathbf M}$ is a {\it symplectic} matrix,
see \cite{A78}, w.~r.~t.~to the standard symplectic structure of $T_{\breve{\mathbf s}}{\mathcal P}$,
but this will not be needed in what follows.\\
We conclude that the eigenvalues and eigenvectors of ${\mathbf M}$ can be obtained as follows:
Let ${\mathbf b}^{(\mu)}$ be the $\mu$-th Fourier basis vector, i.~e.~,
\begin{equation}\label{L3a}
{\mathbf b}^{(\mu)}_\nu =\frac{1}{\sqrt{N}}\exp\left(
\frac{2\pi\,i\,\mu\,\nu}{N}
\right),
\;\mu,\nu=0,\ldots,N-1
\;,
\end{equation}
then we have
\begin{eqnarray}\label{L4a}
&&{\mathbf M}
\left(
\begin{array}{c}
 \sqrt{m^{(\mu,1)}}{\mathbf b}^{(\mu)} \\
 \pm\sqrt{m^{(\mu,2)}}{\mathbf b}^{(\mu)} \\
\end{array}
\right)
=
\left(
\begin{array}{cc}
 0&M^{(1)} \\
 M^{(2)}&0 \\
\end{array}
\right)
\left(
\begin{array}{c}
 \sqrt{m^{(\mu,1)}}{\mathbf b}^{(\mu)} \\
\pm\sqrt{m^{(\mu,2)}}{\mathbf b}^{(\mu)} \\
\end{array}
\right)\\
\label{L4b}
&=&\left(
\begin{array}{c}
 \pm m^{(\mu,1)}\sqrt{m^{(\mu,2)}}{\mathbf b}^{(\mu)} \\
 m^{(\mu,2)}\sqrt{m^{(\mu,1)}}{\mathbf b}^{(\mu)} \\
\end{array}
\right)
=\pm \sqrt{m^{(\mu,1)}m^{(\mu,2)}}
\left(
\begin{array}{c}
 \sqrt{m^{(\mu,1)}}{\mathbf b}^{(\mu)} \\
 \pm\sqrt{m^{(\mu,2)}}{\mathbf b}^{(\mu)} \\
\end{array}
\right)
\;.
\end{eqnarray}
Here the numbers $m^{(\mu,i)},\;i=1,2$ denote the $\mu$-th eigenvalues of
$M^{(i)}$ to be obtained as explained above.
It turns out that all eigenvalues $m^{(\mu,1)}$ are positive and all $m^{(\mu,2)}$
negative, hence the eigenvalues in (\ref{L4b}) will always be of the form $\pm {\sf i}\, \omega_\mu,\;\omega_\mu>0$,
as it should be. The corresponding wave numbers are $k_\mu\equiv \frac{2\pi\mu}{N},\;\mu=0,\ldots,N-1$ which are usually
chosen to cover the interval $(-\pi,\pi)$ (note that $\mu$ is only defined modulo $N$ and hence could also be chosen to run
from $-\lfloor \frac{N}{2}\rfloor$ to $\lfloor \frac{N}{2}\rfloor$). The set of pairs
$(k_\mu,\omega_\mu),\;\mu=-\lfloor \frac{N}{2}\rfloor,\ldots,\lfloor \frac{N}{2}\rfloor$
constitutes the finite ``dispersion relation" for the corresponding dipole ring.
It remains to give explicit
expressions for the first row of the sub-matrices $M^{(i)}$. After some algebra
we obtained the following result
\begin{eqnarray}\label{L5a}
M^{(1)}_{1,k}&=&
\left\{
\begin{array}{l@{\;:\;}l}
 \frac{1}{8}\sum_{j=1}^{N-1}
\left(
2\csc^3 \frac{\pi j}{N}-\csc \frac{\pi j}{N}
\right)& k=0\,,\\
\frac{1}{8}\csc^3 \frac{\pi k}{N}
 & k=1,\ldots,N-1,
\end{array}
\right.\\
\label{L5b}
M^{(2)}_{1,k}&=&
\left\{
\begin{array}{l@{\;:\;}l}
 \frac{1}{8}\sum_{j=1}^{N-1}
\left(
-2\csc^3 \frac{\pi j}{N}+\csc \frac{\pi j}{N}
\right)& k=0\,,\\
-\frac{1}{8}\left(
\csc^3 \frac{\pi k}{N}+\csc \frac{\pi k}{N}
\right)
 & 1,\ldots,N-1.
\end{array}
\right.
\end{eqnarray}

For the sake of illustration we consider the case $N=3$. The sub-matrices $M^{(1)}$ and $M^{(2)}$ assume the form
\begin{equation}\label{L6}
M^{(1)}=\frac{1}{6\sqrt{3}}\left(
\begin{array}{ccc}
 5&2&2\\
 2&5&2\\
 2&2&5\\
\end{array}
\right)
\;,
\end{equation}
and
\begin{equation}\label{L6}
M^{(2)}=\frac{1}{12\sqrt{3}}\left(
\begin{array}{ccc}
 -10&-7&-7\\
 -7&-10&-7\\
 -7&-7&-10\\
\end{array}
\right)
\;.
\end{equation}
Their eigenvalues are $\sqrt{3}$ times the Fourier series of the first row of these matrices, which gives
\begin{equation}\label{L7}
m^{(1)}=\left(\frac{\sqrt{3}}{2},\frac{1}{2\sqrt{3}},\frac{1}{2\sqrt{3}}\right),\quad
m^{(2)}=\left(-\frac{2}{\sqrt{3}},-\frac{1}{4\sqrt{3}},-\frac{1}{4\sqrt{3}}\right)
\;.
\end{equation}
The square root of the respective products gives the eigenvalues  $\pm {\sf i}\,\omega_\mu$ of $M$ where
$\omega_0=1,\;\omega_{\pm 1}=\frac{1}{2\sqrt{6}}$ corresponding to the wave numbers $k_0=0$ and $k_{\pm 1}=\pm \frac{2\pi}{3}$,
respectively.

The analogous results for the finite dispersion relations of dipole rings with $N=4,\ldots,7$ are given in table \ref{Tab1}.
The set of pairs $(\frac{\omega_\mu}{\omega_0},k_\mu),\; \mu=0,\ldots,\lfloor\frac{N}{2}\rfloor$ will be called the ``normalized"
dispersion relation. It is displayed in Fig.~\ref{fig_dri} for $N=3,\ldots,300$ and can be seen there to approach the dispersion relation
of the infinite dipole chain.\\

\begin{table}
\caption{{\label{Tab1}}Table of spin wave frequencies for $N=3,\ldots,7$ obtained by analytical methods. The explicit analytical
formulas for $N=7$ are too lengthy to be listened in the table.\\}
\begin{center}
\begin{tabular}{|c|c|c|c|c|}\hline
$N$ & $\omega_0$ & $\omega_1$ & $\omega_2$ & $\omega_3$ \\ \hline\hline
$3$ & $1$ & $\frac{1}{2\sqrt{6}}$ & & \\ \hline
 & $1.0$ & $0.204124$ &&\\ \hline
 $4$ & $\frac{1}{4} \sqrt{\frac{3}{2} \left(41+9 \sqrt{2}\right)}$ & $\frac{1}{4} \sqrt{18-\frac{3}{\sqrt{2}}}$ &
 $\frac{1}{4} \sqrt{\frac{3}{2} \left(1+\sqrt{2}\right)}$ & \\ \hline
  & $2.24432$ & $0.996202$ & $0.475744$ &\\ \hline
  $5$ & $\sqrt{\frac{63}{5}+\frac{57}{5 \sqrt{5}}}$ & $\frac{1}{10} \sqrt{615-\frac{9 \sqrt{5}}{2}}$ &
   $\sqrt{\frac{37}{40}+\frac{71}{40 \sqrt{5}}}$ & \\ \hline
    & $4.20693$ & $2.45955$ & $1.31103$ &\\ \hline
    $6$ & $\sqrt{\frac{1205}{32}+\frac{337}{16 \sqrt{3}}}$
    & $\frac{1}{4} \sqrt{\frac{11}{3} \left(91+5 \sqrt{3}\right)}$
    & $\sqrt{\frac{697}{96}+\frac{2}{\sqrt{3}}}$
    & $\frac{1}{4} \sqrt{37+23 \sqrt{3}}$ \\ \hline
    & $7.05809$ & $4.779$ & $2.90088$ & $2.19142$ \\ \hline
    $7$ & $\ast$& $\ast$& $\ast$& $\ast$\\ \hline
    & $10.9695$
     & $8.13214$
     & $5.44092$
     & $3.87036$ \\ \hline
\end{tabular}
\end{center}
\end{table}

\section{Numerical results}\label{sec:S}

\begin{figure}
\begin{center}
\includegraphics[clip=on,width=150mm,angle=0]{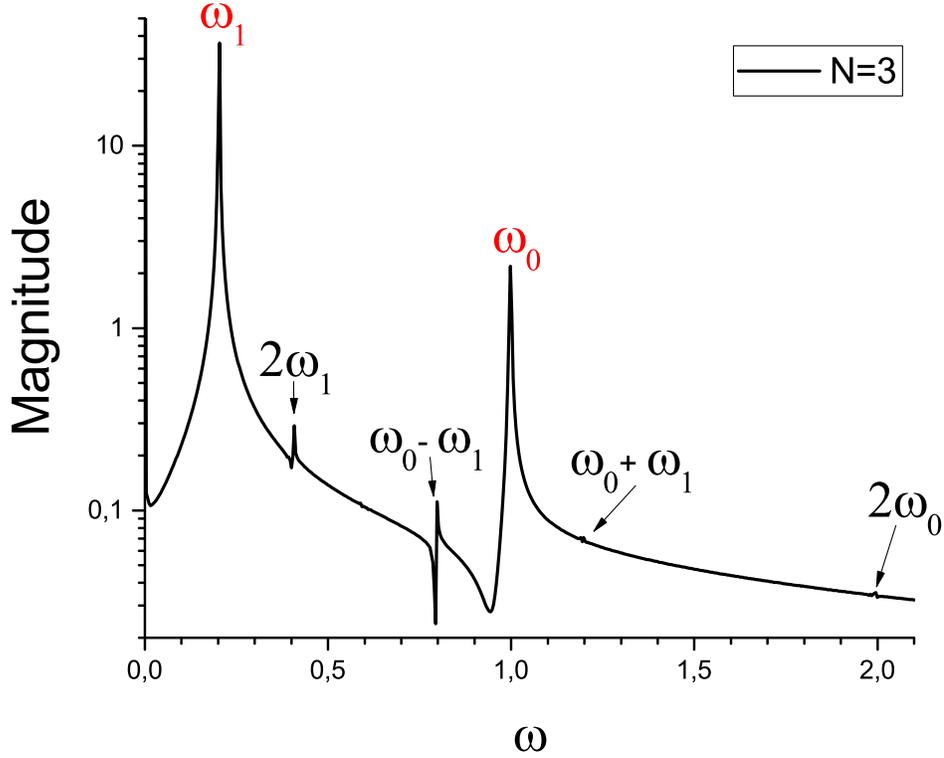}
\end{center}
\caption{Fourier time transform of the autocorrelation
function $\langle ac(t)\rangle$ vs.~$\omega$ for a dimensionless
temperature of $T=0.001$ showing two main peaks at
$\omega_1 = 0.2041$ and $\omega_0 = 0.9976$.
Using this semilog scale the combined frequencies can be identified as well.
\label{fig_AC3spin}
}
\end{figure}

\begin{figure}
\centering
	\includegraphics[width=0.4\linewidth]{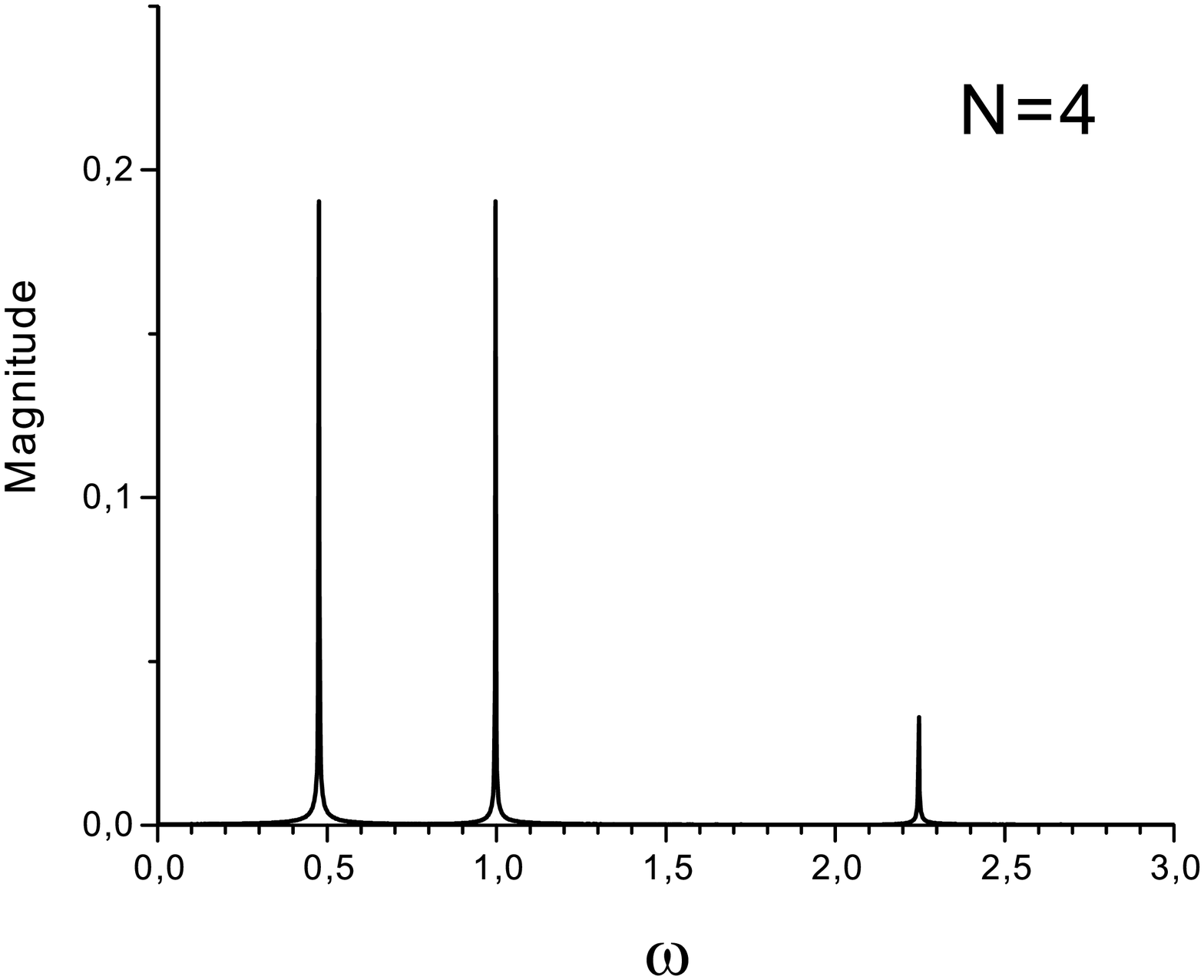}
	\includegraphics[width=0.4\linewidth]{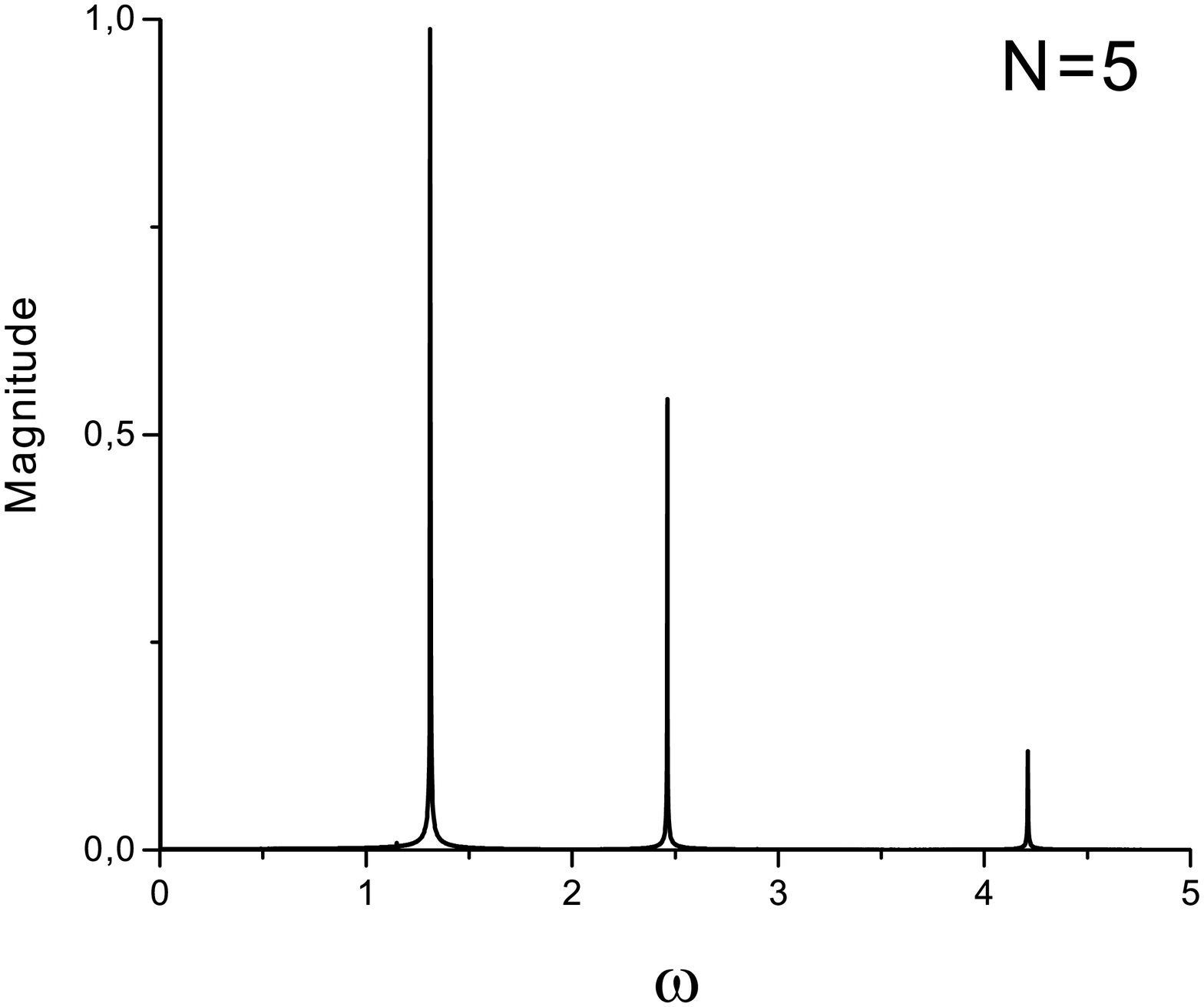}
	\includegraphics[width=0.4\linewidth]{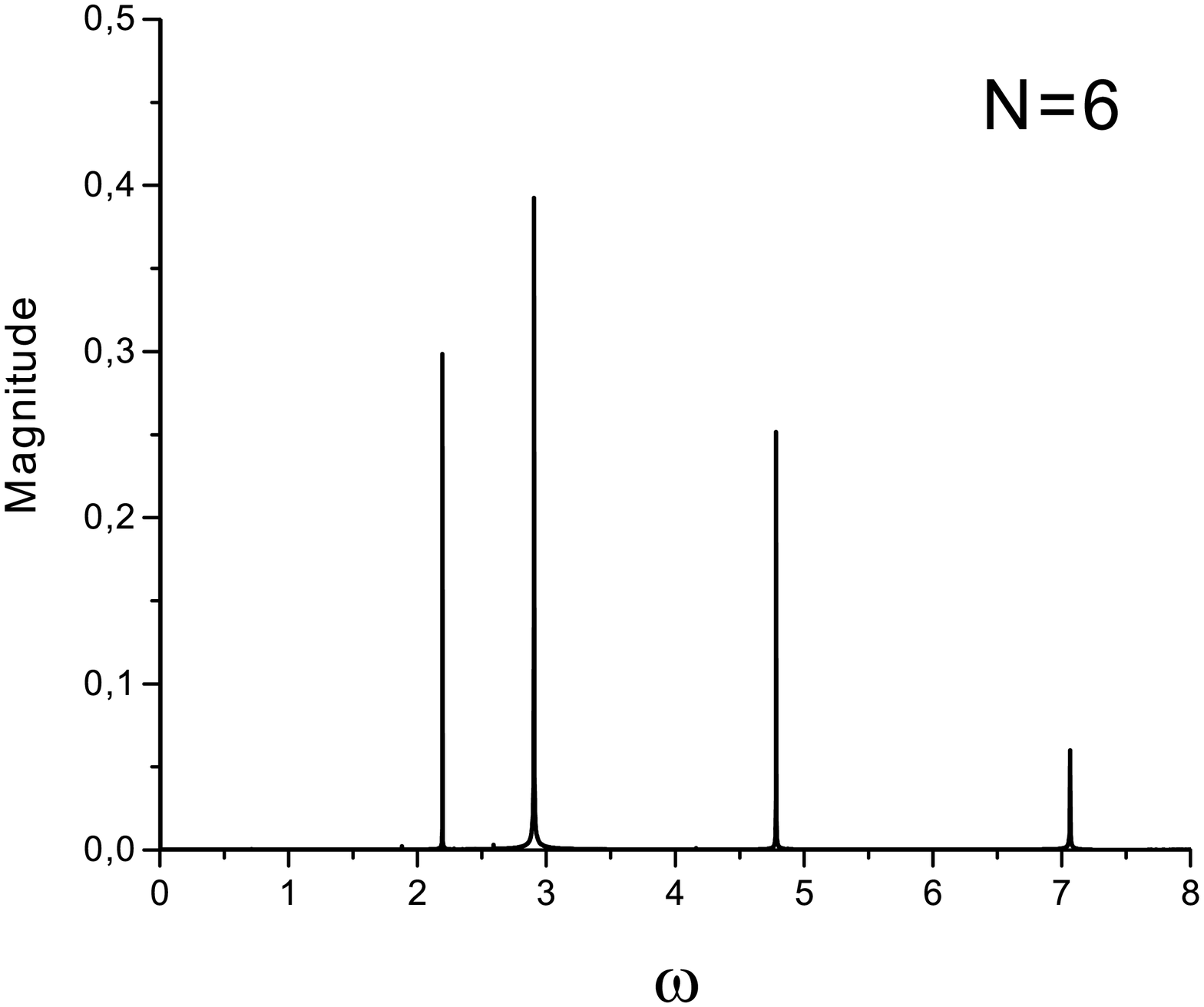}
	\includegraphics[width=0.4\linewidth]{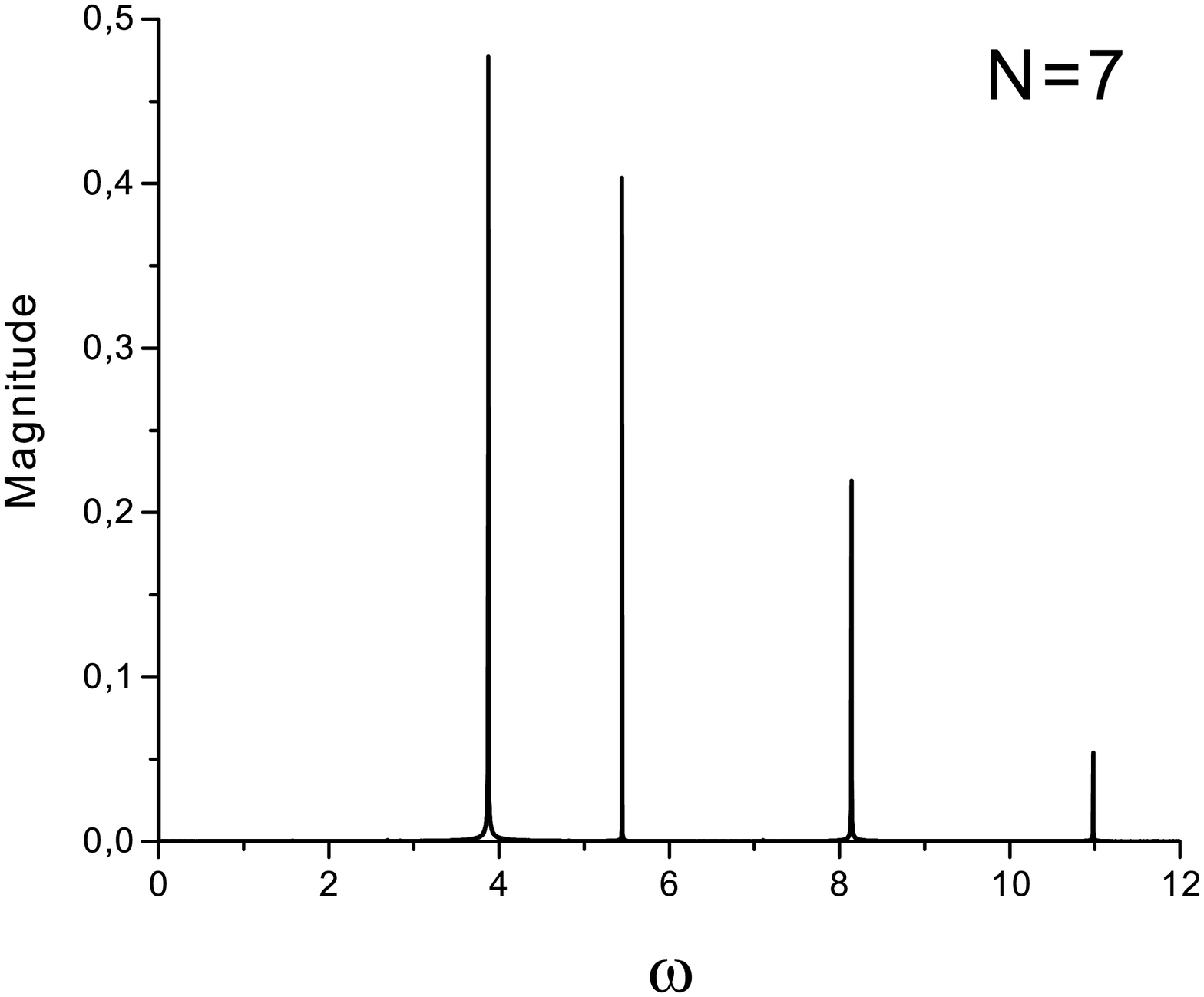}
\caption{Fourier time transform of the autocorrelation
function $\langle ac(t)\rangle$ vs.~$\omega$ for a dimensionless
temperature of $T=0.001$ for rings containing 4 to 7 dipoles. The peak positions coincide with the predicted ones,
see table \ref{Tab1},
 within an accuracy of $0.25\%$.
\label{fig_ACspinrings}
}
\end{figure}

\begin{figure}
\begin{center}
\includegraphics[clip=on,width=150mm,angle=0]{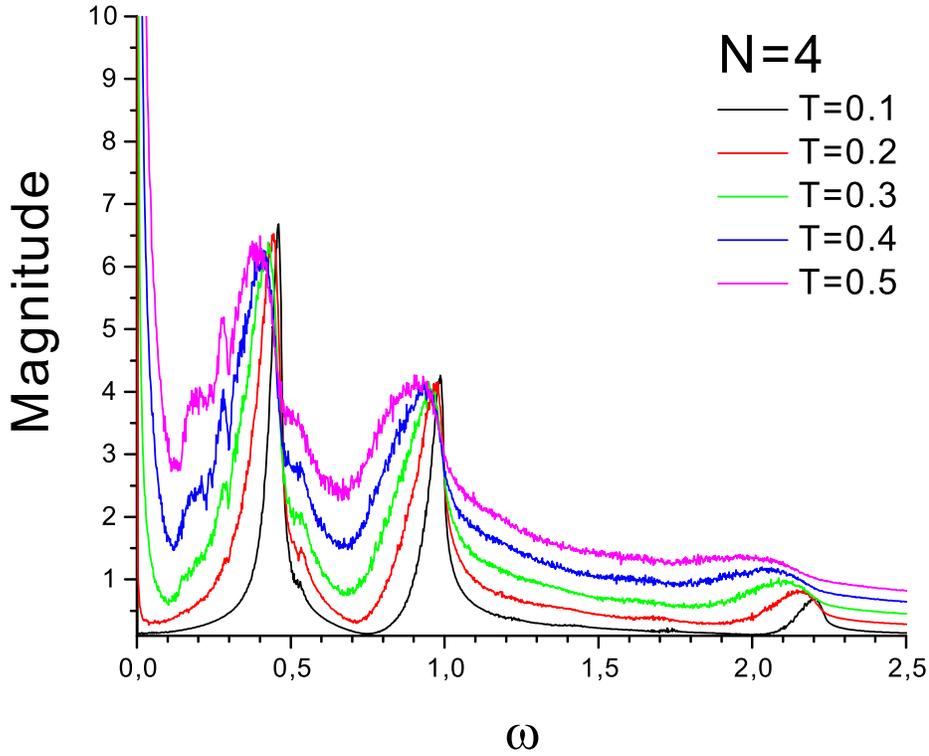}
\end{center}
\caption{Fourier transform of the autocorrelation
function $\langle ac(t)\rangle$ vs.~$\omega$ for the case $N=4$ at dimensionless
temperatures between $T=0.1$ and $T=0.5$.
\label{fig_AC4spinT}
}
\end{figure}

\begin{figure}
\centering
	\includegraphics[width=0.4\linewidth]{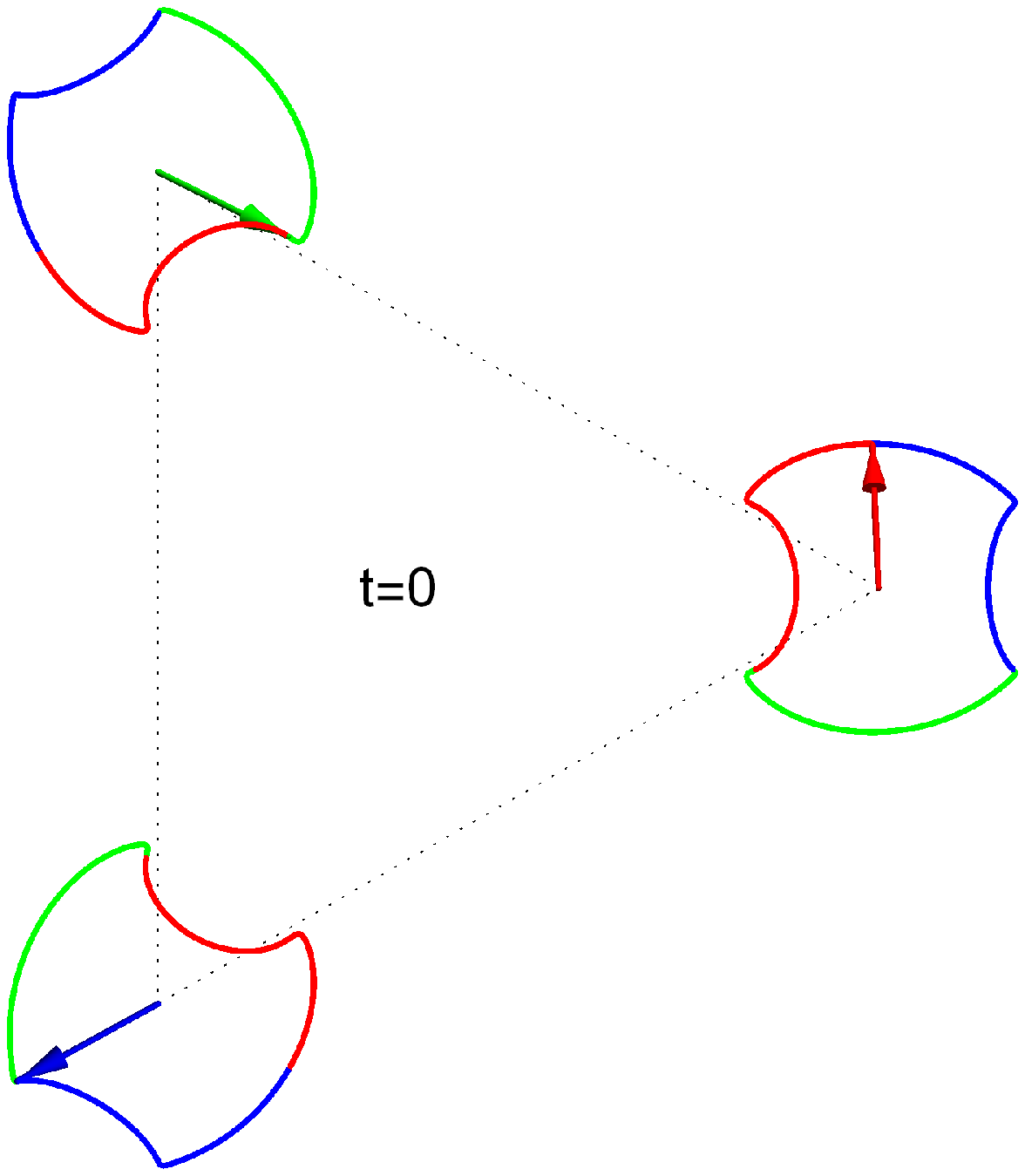}
	\includegraphics[width=0.4\linewidth]{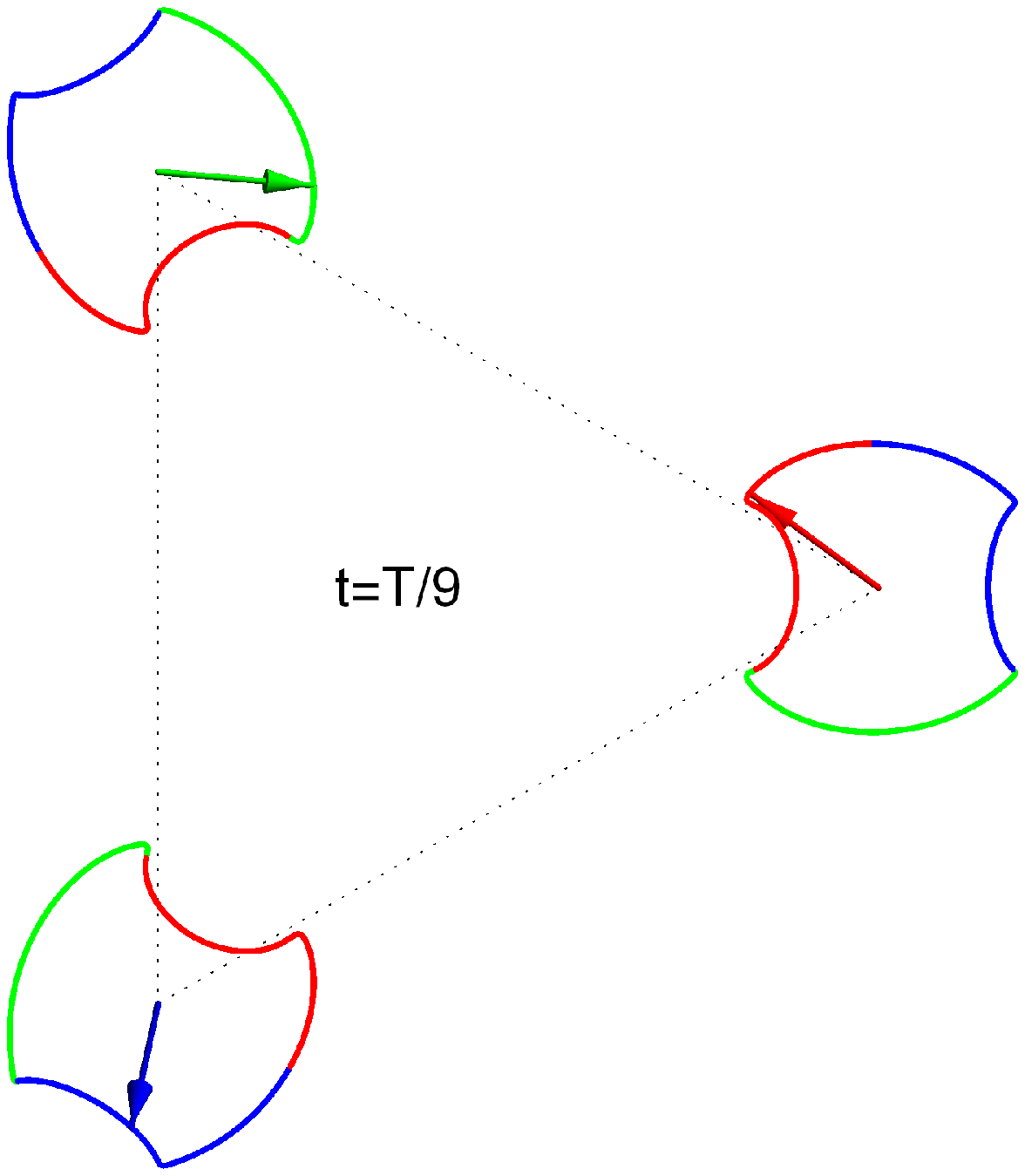}
	\includegraphics[width=0.4\linewidth]{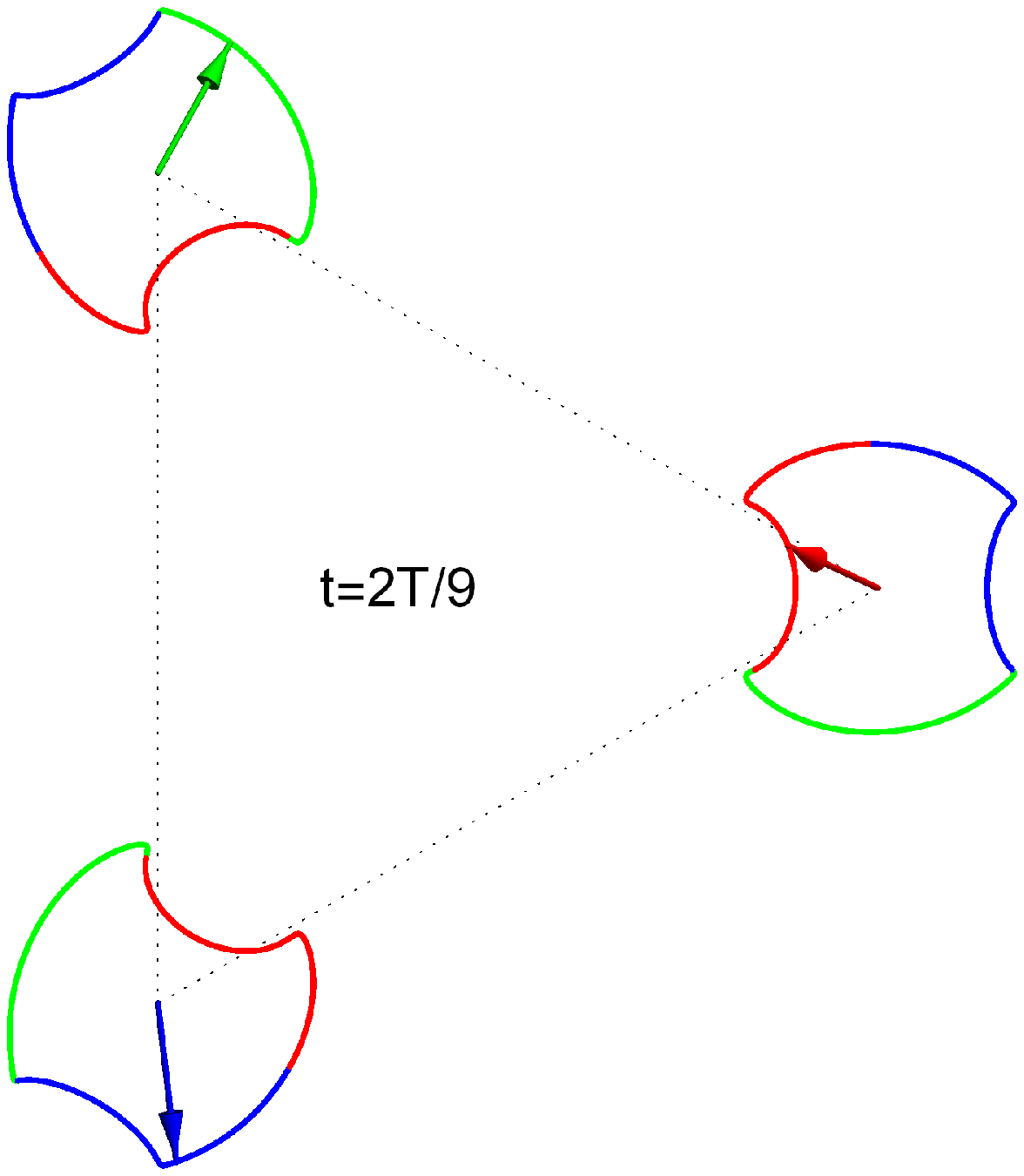}
	\includegraphics[width=0.4\linewidth]{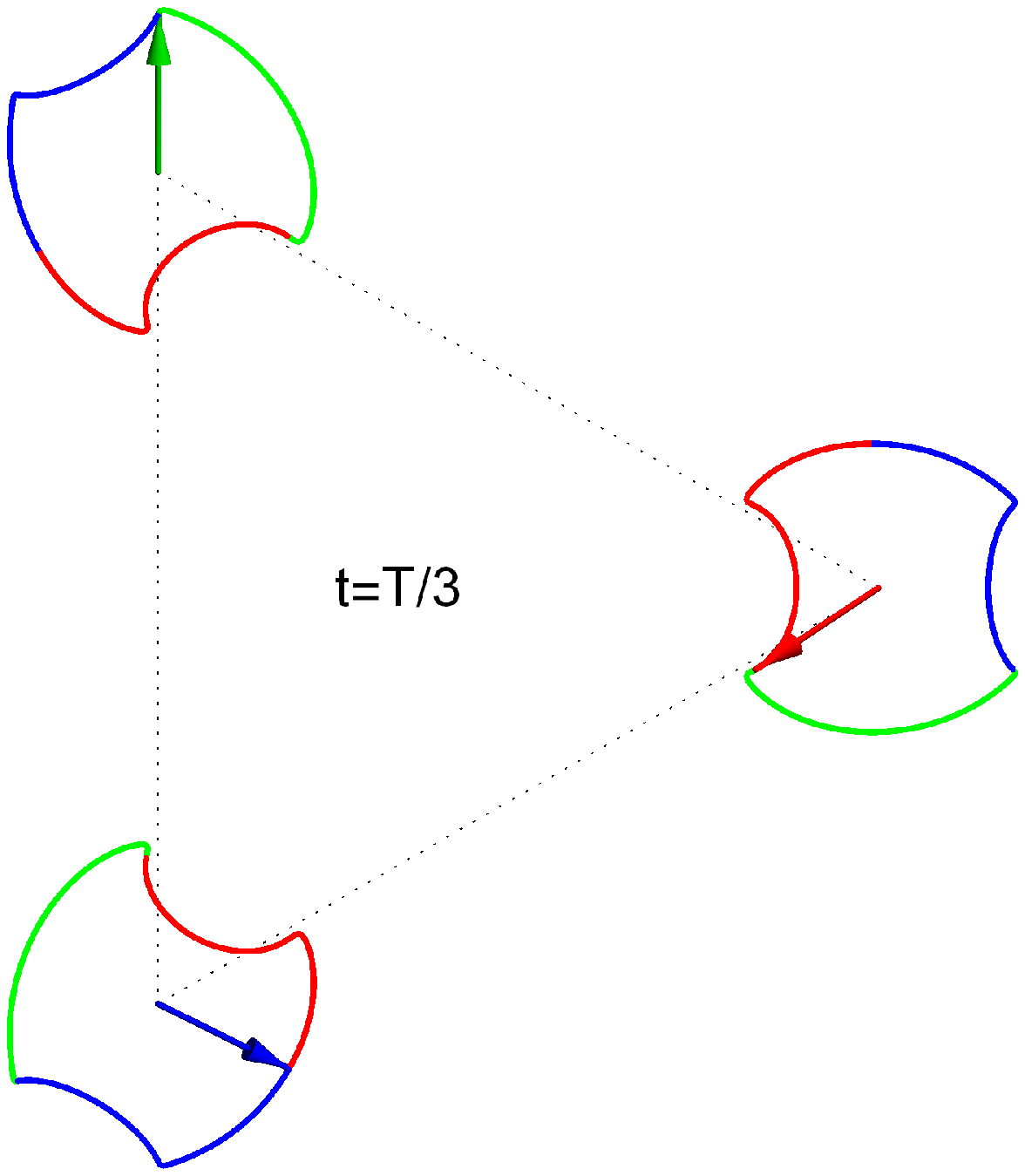}
\caption{Three-dimensional plot of the $N=3$ spin wave with initial conditions according to table \ref{Tab2}. The three colors (red, green, blue) correspond to the three spins $\mu=0,1,2$.
The three solutions are only plotted for $t=0,\frac{T}{9},\frac{2T}{9},\frac{T}{3}$, but according to the property of ``choreography"
(\ref{F1}) their orbits together form the closed curve that is periodically traversed by each spin vector.
\label{fig_cognac}
}
\end{figure}

\subsection{Method}
A direct experimental test of the results of the previous sections for real systems is naturally affected by thermal fluctuations due to finite temperatures. Hence it seems worthwhile to investigate the thermodynamics of the dipolar spin waves, especially to determine the temperature dependent equilibrium autocorrelation function ($ac$). In order to calculate the canonical ensemble average numerically we used the so-called "Gibbs approach" \cite{LL99}, where the trajectories ${\mathbf s}_\mu(t)$ for the dipoles are calculated for the \textit{isolated} system by solving the equations of motion (\ref{R2}) over a certain number of time steps numerically.
The initial conditions for each trajectory are generated by a standard Monte Carlo simulation for a temperature $T$. By averaging all generated trajectories at each time step one obtains the canonical ensemble average.

\subsection{Autocorrelation functions}
We define the thermal equilibrium autocorrelation function, to be denoted by $\langle ac \rangle(t,T)$, as the canonical
ensemble average of ${\mathbf s}_1(0)\cdot {\mathbf s}_1(t)$, namely  $\langle ac \rangle(t,T)=\langle{\mathbf s}_1(0)\cdot {\mathbf s}_1(t)\rangle_T$.
Due to the symmetry of the system the results for all dipoles are identical. In Fig.~\ref{fig_AC3spin} we show the Fourier time transform
of $\langle ac \rangle(t,T)$ calculated using Monte Carlo methods for an equilateral triangle of interacting dipoles for an dimensionless temperature
of $T=0.001$. We expect that the peaks of the transform should approach the frequencies of the spin waves or the combined frequencies
in the limit $T\longrightarrow 0$. Indeed, the spectrum shows two main peaks at the predicted theoretical frequencies
 $\omega_1 = \frac{1}{2\sqrt{6}}$ and $\omega_0 = 1$.
 Furthermore, the theoretically expected combined frequencies
 $2\omega_0, 2\omega_1$, $\omega_1 \pm \omega_0$, and $\omega_0 \pm \omega_1$
 are visible in this semilog plot as well, however with a much lower magnitude, i.~e.~invisible on a linear scale.
As in the case $N=2$, see \cite{SSHL15}, we expect that the peaks at the combined frequencies are suppressed due to the process of thermal averaging.

In order to illustrate the changes in the spectrum for increasing ring sizes we have simulated the $\langle ac \rangle$ for a variety of small ring systems, i.e. for $N=4, 5, 6, 7$ coupled dipoles (see Fig.~\ref{fig_ACspinrings}). All results fully agree with our theoretical predictions
for the limit $T\longrightarrow 0$.
According to table \ref{Tab1} we find three peaks for $N=4$ and $N=5$ dipoles and four peaks for $N=6$ and $N=7$ dipoles.

Another important aspect with respect to any experimental test of our theory is the temperature dependence of the $\langle ac \rangle$. We have performed simulations for the case of $N=4$ for 5 different temperatures. The results are given in Fig.~\ref{fig_AC4spinT}. One can see that the peaks become broader and eventually wash out with increasing temperatures. The peak positions are shifted towards lower frequencies with increasing temperatures.

\subsection{Spin waves with finite amplitudes for $N=3$}
\label{sec:F}
By numerically adapting the initial conditions we have obtained a rich variety of finite amplitude solutions of the eom that are very close to
exact spin waves.
These solutions will be investigated elsewhere; in this article we will only provide an example far from the linear regime.
Concentrating on the simplest case $N=3$ we have transformed the eom to the set of coordinates
${\mathbf a}_\mu\equiv(\xi_\mu,\eta_\mu,\zeta_\mu),\;\mu=0,1,2$
defined in (\ref{SO1}). This has the consequence that spin waves with wave number $k_1=2\pi/3$ can be characterized by the property of ``choreography", i.~e.~,
\begin{equation}\label{F1}
{\mathbf a}_\mu(t+T/3)={\mathbf a}_{\mu+1}(t),\;
\mu=0,1,2
\;,
\end{equation}
where $T$ is the period of any single spin solution and the addition $\mu+1$ is understood modulo $3$. It says that all spin vectors
traverse the same closed curve but with a mutual phase shift of $T/3$.
With the initial conditions given in table \ref{Tab2} we obtained the numerical solution displayed in Fig.~\ref{fig_cognac}.

\begin{table}
\caption{{\label{Tab2}}Table of initial conditions for an $N=3$ spin wave using the coordinates (\ref{SO1}).
We provide also the numerical values for the period $T$ and the energy $E$.\\}
\begin{center}
\begin{tabular}{lll}\hline
$\xi_0=0$ &$\eta_0= 0.916737$ &$\zeta_0=  -0.399492$ \\
$\xi_1= -0.791657$ & $\eta_1=  -0.502101$ &$\zeta_1=  0.348101$ \\
$\xi_2=-\xi_1$ &$\eta_2= \eta_1$  & $\zeta_2= \zeta_1$ \\ \hline
$T=38.0019 $ & $E=0.10674 $& \\ \hline
\end{tabular}
\end{center}
\end{table}

\section{Synchronized oscillations}
\label{sec:SO}

\begin{figure}
\begin{center}
\includegraphics[clip=on,width=150mm,angle=0]{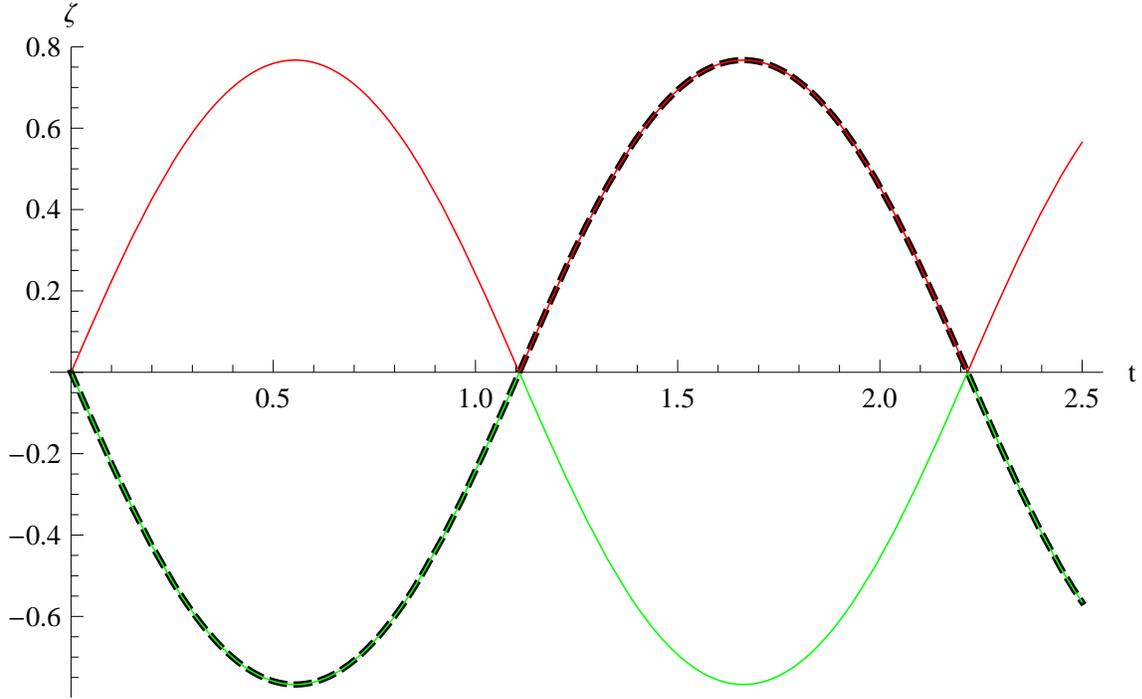}
\end{center}
\caption{Numerical solution $\zeta(t)$ of the eom (\ref{R2})
for $N=5$ and initial conditions (\ref{SO11}) adapted to synchronized oscillations (black dashed curve).
This solution locally coincides with one of the the analytical solutions (\ref{SO10c}) (red and green curves).
\label{fig_so}
}
\end{figure}

\begin{figure}
\begin{center}
\includegraphics[clip=on,width=150mm,angle=0]{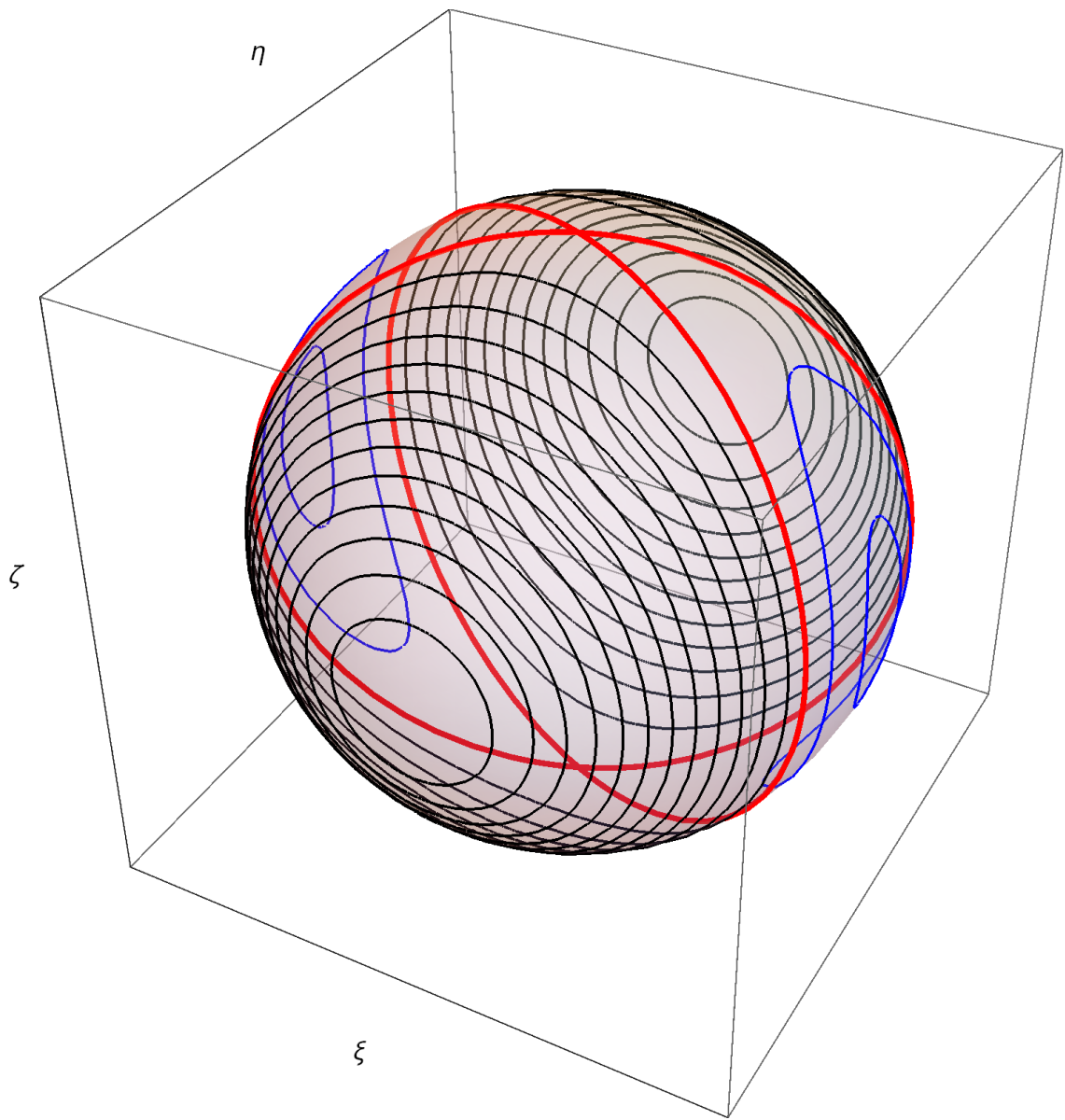}
\end{center}
\caption{Plot of different orbits of synchronized oscillations according to the eom (\ref{SO2}) for $N=5$.
There are two regions of stable oscillations around the states with minimal and maximal energy (black and blue curves, resp.~) separated
by orbits that represent ``switch solutions" (red curves).
\label{fig_sp}
}
\end{figure}

\begin{figure}
\begin{center}
\includegraphics[clip=on,width=150mm,angle=0]{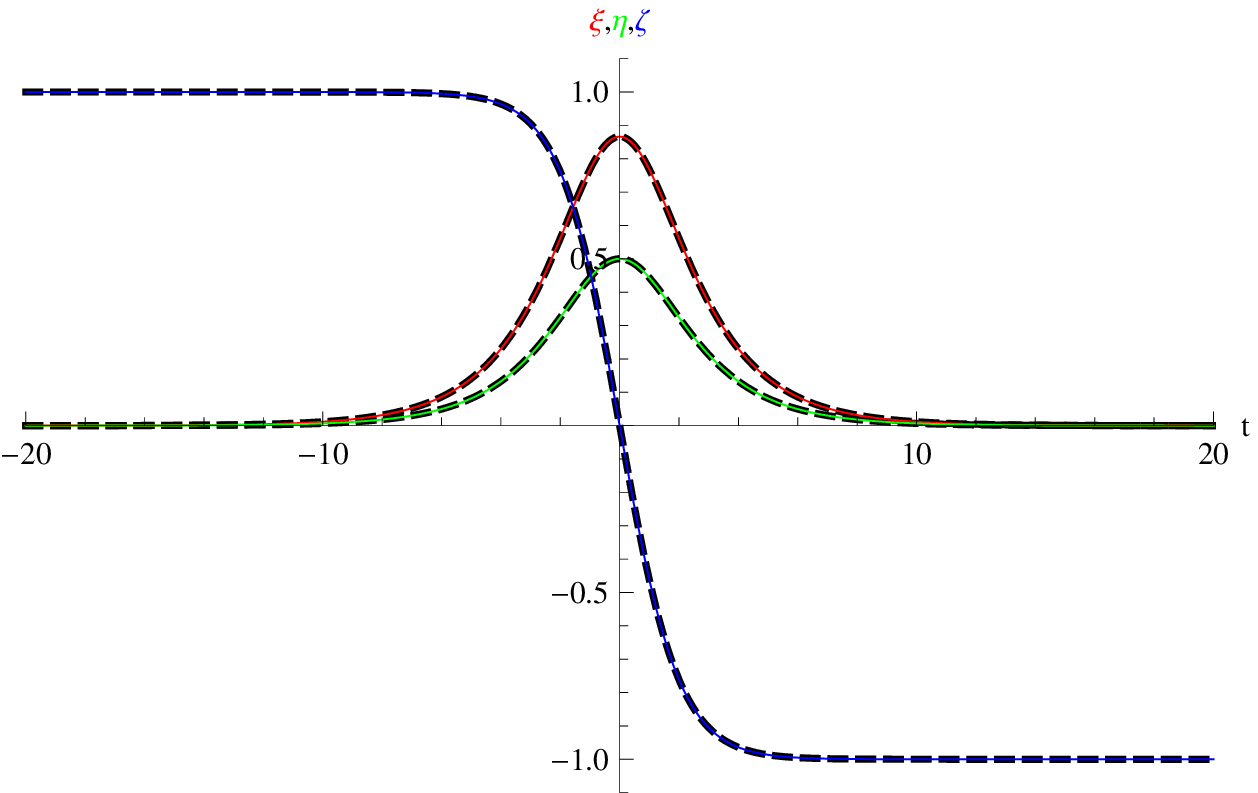}
\end{center}
\caption{Plot of a numerical ``switch solution" as a degenerate case of synchronized oscillations according to the eom (\ref{SO2}) for $N=3$
(black dashed curves) and the corresponding analytical solution (\ref{SO12}) (colored curves).
\label{fig_sw}
}
\end{figure}

There exists a family of analytical solutions of the eom that can be understood as
spin waves with $k=0$ and infinite phase velocity. Each moment vector performs a periodic
non-linear oscillation and there is no phase shift between these oscillations. Moreover, the
collective oscillations are all the same if described in the local coordinate system adapted to the ground
state $\breve{\mathbf s}$, see (\ref{R5}).

More specifically, we introduce coordinates
\begin{equation}\label{SO1}
\xi_\mu={\mathbf r}_\mu\cdot{\mathbf s}_\mu,\;\eta_\mu={\breve{\mathbf s}}_\mu\cdot{\mathbf s}_\mu,\;,
\zeta_\mu={\mathbf e}_3\cdot{\mathbf s}_\mu,\quad \mu=0,\ldots,N-1\;,
\end{equation}
and assume that for all moment vectors ${\mathbf s}_\mu$ these coordinates are the same and will be simply denoted by $(\xi,\eta,\zeta)$.
Mathematically, we consider the sub-manifold ${\mathcal Q}$ of the phase space ${\mathcal P}$
consisting of all fixed points of the shift operator $S$, i.~e.~,
${\mathcal Q}\equiv \{{\mathbf s}\in{\mathcal P}\,|\,S\,{\mathbf s}={\mathbf s}\}$.
This sub-manifold is invariant under the time evolution $T_t$ since
$S\,{\mathbf s}={\mathbf s}$ implies $S\left(T_t\,{\mathbf s}\right)=T_t\,S\,{\mathbf s}=T_t\,{\mathbf s}$
where the commutation relation (\ref{R7}) has been used.

Then it is obvious that the eom will assume the simple form
\begin{equation}\label{SO2}
\frac{d}{dt}\xi=\alpha\;\eta\,\zeta,\quad
\frac{d}{dt}\eta=\beta\;\zeta\,\xi, \quad
\frac{d}{dt}\zeta=\gamma\;\xi\,\eta
\;,
\end{equation}
where the three real constants $\alpha,\beta,\gamma$ have still to be determined.
They satisfy two constraints due to the conserved quantities
\begin{eqnarray}\label{SO3a}
\xi^2+\eta^2+\zeta^2&=&1\;,\\
\label{SO3b}
H=a\,\xi^2+b\,\eta^2+d\,\zeta^2&=&\epsilon
\;.
\end{eqnarray}
The constants $a,b,d$ can be obtained as $N$ times the constant row sum of certain matrices
$A,B,D$, see \ref{sec:A}, but for the moment we prefer to work with the $a,b,d$ as undetermined constants.
The special solution $\xi=0,\;\eta=1,\;\zeta=0$ of (\ref{SO2})
represents the ground state $\breve{\mathbf s}$ that is a stationary point of the eom.

We notice that the eom (\ref{SO2}) are of the same form as Euler's equation describing the time evolution of the angular momentum
${\mathbf M}$
of a rigid body in the body frame, see \cite{A78}. In that case the equation analogous to (\ref{SO3b}) reads
\begin{equation}\label{SO4}
2\,E\,=\,\frac{M_1^2}{I_1}+\,\frac{M_2^2}{I_2}+\,\frac{M_3^2}{I_3}
\;,
\end{equation}
where the $I_1,I_2,I_3>0$ are the body's principal moments of inertia. In our case, however, the analogous coefficients
$a,b,d$ have different signs, namely $a,d>0$ but $b<0$, see \ref{sec:A}. Hence the analogy is not complete; nevertheless
the mathematical treatment of (\ref{SO2}) will be very similar to the case of Euler's equation. In particular, the well-known
facts about the stability of the rigid body's rotations about the two principal axes of inertia corresponding to the maximal and the
minimal moment of inertia apply to the collective oscillations of the dipole ring as well, see Fig.~\ref{fig_sp}.

We now turn to the eom for the function $\zeta(t)$, see (\ref{SO2}). After some algebra we obtain for the corresponding parameter
\begin{equation}\label{SO5}
\gamma = -\frac{3}{8} \sum _{\mu =1}^{N-1} \csc^3 \left(\frac{\pi  \mu }{N}\right)
\;.
\end{equation}
The two conserved quantities (\ref{SO3a}),(\ref{SO3b}) can be used to express $\xi$ and $\eta$ through $\zeta$:
\begin{eqnarray}\label{SO6a}
\xi&=&\pm \frac{\sqrt{\zeta ^2 (b-d)-b+\epsilon }}{\sqrt{a-b}}\;,\\
\label{SO6b}
\eta&=&\pm \frac {\sqrt {\zeta ^2 (d - a) + a - \epsilon }} {\sqrt {a - b}}
\;.
\end{eqnarray}
Hence the analytical solution for $\zeta(t)$ suffices to solve the system (\ref{SO2}) completely.

Separation of variables transforms the $3$rd eq.~of (\ref{SO2}) into an elliptic integral:
\begin{eqnarray}\label{SO7a}
\int\,dt&=&\int\frac{d\zeta}{\gamma\,\xi\,\eta}=\int\frac{a-b}{\gamma  \sqrt{(a-d) (d-b)}
 \sqrt{\frac{a-\epsilon }{a-d}-\zeta ^2} \sqrt{\frac{b-\epsilon}{b-d}-\zeta ^2}}\,d\zeta\\
\nonumber
&=& \frac{1}{\lambda}\int \frac{d\zeta}{\sqrt{(f+\zeta^2)(g+\zeta^2)}}
=
\frac{1}{\lambda}\int \frac{dv}{\sqrt{4 v^3-g_2 v-g_3}}
\equiv \frac{1}{\lambda}\int \frac{dv}{\sqrt{P(v)}}
\;.\\
\label{SO7b}
\end{eqnarray}
The first equality in (\ref{SO7b}) follows from the definitions
\begin{equation}\label{SOsub}
f=-\frac{a-\epsilon}{a-d},\quad g=-\frac{b-\epsilon}{b-d},\quad\lambda=\frac{\gamma\sqrt{(a-d)(b-d)}}{a-b}
\;,
\end{equation}
and the second one from the substitution
\begin{equation}\label{SO8}
v=\zeta^2-\frac{f+g}{3}\equiv \zeta^2+v_0
\end{equation}
and the definitions
\begin{eqnarray}\label{SO9a}
g_2&=&\frac{4}{3} \left(f^2-f g+g^2\right),\\
\label{SO8b}
g_3&=&\frac{4}{27} (f-2 g) (2 f-g) (f+g)
\;.
\end{eqnarray}
By inserting appropriate boundaries and using the definition of the Weierstrass elliptic
function ${\mathcal P}(z)\equiv{\mathcal P}(z;g_2,g_3)$, see \cite{AS72} Ch.~$18$, we obtain
\begin{eqnarray}\label{SO10a}
\lambda\,t&=&\int_{v_0}^v\, \frac{dv}{\sqrt{P(v)}}
=\int_{\infty}^v\, \frac{dv}{\sqrt{P(v)}}-\int_{\infty}^{v_0}\, \frac{dv}{\sqrt{P(v)}}
\equiv u-u_0,\\
\label{SO10b}
&&v={\mathcal P}(u),\quad v_0={\mathcal P}(u_0),\\
\label{SO10c}
\zeta(t)&=&\pm\sqrt{
{\mathcal P}
(\lambda t+u_0)-v_0}
\;.
\end{eqnarray}

As an example we consider the case $N=5$ and initial conditions
\begin{equation}\label{SO11}
\xi(0)=\eta(0)=\frac{1}{2}\sqrt{2},\quad
\zeta(0)=0
\;.
\end{equation}
The result of the numerical integration of the eom compared with the analytical solution is shown in Fig.~\ref{fig_so}.
This solution belongs to the oscillations about the system's ground state (black curves in Fig.~\ref{fig_sp}).
There exist similar oscillations about the state with maximal energy (blue curves in Fig.~\ref{fig_sp}). Both regions
of stable oscillations are separated by orbits that represent transitions between the two unstable stationary states
$\xi=\eta=0,\;\zeta=\pm 1$ (red curves in Fig.~\ref{fig_sp}). These ``switch solutions" can be expressed in terms
of elementary functions, but they are of increasing complexity with growing $N$. Here we only present the simplest case of $N=3$:
\begin{equation}\label{SO12}
\left(
\begin{array}{c}
 \xi(t) \\
 \eta(t)\\
\zeta(t)\\
\end{array}
\right)
=
\left(
\begin{array}{c}
 \frac{1}{2} \sqrt{3}\, \mbox{sech}\left(\frac{t}{2}\right) \\
 \frac{1}{2}\, \mbox{sech}\left(\frac{t}{2}\right) \\
 -\tanh \left(\frac{t}{2}\right) \\
\end{array}
\right)\;,
\end{equation}
see Fig.~\ref{fig_sw}.

\section{Spin waves of the infinite dipole chain}
\label{sec:DC}

\begin{figure}
\begin{center}
\includegraphics[clip=on,width=150mm,angle=0]{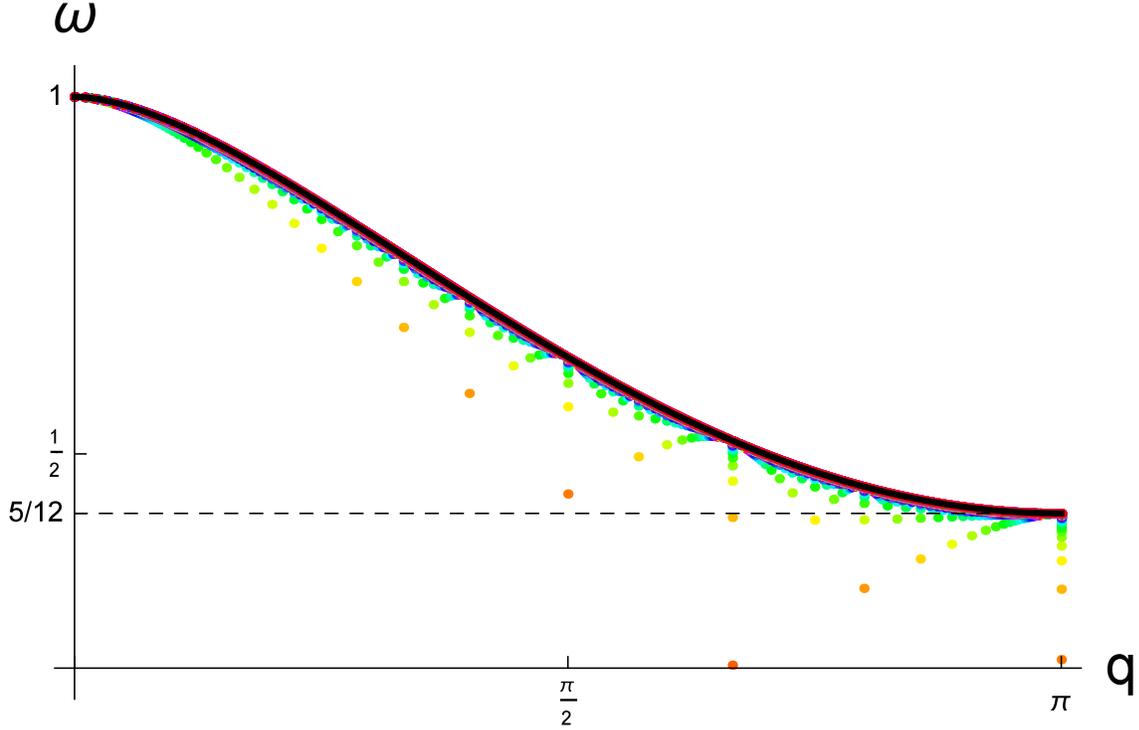}
\end{center}
\caption{Comparison of the dispersion relation $\omega=\omega(q,z)$ according to (\ref{DC6b}) with $z=1$ for the infinite dipole chain
(black curve)
with the normalized dispersion relations for finite dipole rings (colored points) for $N=3,\ldots,300$. The dashed line indicates the
value $\omega(\pi,1)=\frac{5}{12}$. It is evident that the normalized dispersion relations approach the graph of $\omega=\omega(q,1)$ for $N\longrightarrow\infty$.
\label{fig_dri}
}
\end{figure}

The dipoles will be indexed by the set of integers ${\mathbb Z}$
and placed on the z-axis with distance $a$  between adjacent neighbors.
Hence ${\mathbf e}_3$ is the unit vector in the direction joining any two dipoles. The constant $a$  is absorbed into the choice of the unit of time as
in the case of the dipole ring. Hence the dimensionless equation of motion for the dipole with index
$m\in{\mathbb Z}$ reads
\begin{equation}\label{DC1}
\frac{d}{dt}{\mathbf s}_m = {\mathbf s}_m \times \sum_{n\in{\mathbf Z},\,n\neq m}\frac{1}{|n-m|^3}
\left( {\mathbf s}_n-3\,{\mathbf s}_n\cdot {\mathbf e}_3\;{\mathbf e}_3
\right)
\;.
\end{equation}
We make the following ansatz for a spin wave:
\begin{equation}\label{DC2}
{\mathbf s}_n(t)=
\left(
\begin{array}{c}
 \sqrt{1-z^2} \cos (n \,q- \omega\,t ) \\
 \sqrt{1-z^2} \sin (n \,q- \omega\,t ) \\
 z \\
\end{array}
\right),\quad n\in{\mathbb Z}
\;,
\end{equation}
where $\omega,q$ and $z$ are parameters, constant in time, to be determined later.
It will suffice to check whether (\ref{DC2}) satisfies (\ref{DC1}) for the special
case $m=0$ and $t=0$ since the general case is analogous. Hence we have to show that
the second term of the r.~h.~s.~of (\ref{DC1}) is of the form
\begin{equation}\label{DC3}
\sum_{n\in{\mathbf Z},\,n\neq 0}\frac{1}{|n|^3}
\left( {\mathbf s}_n-3\,{\mathbf s}_n\cdot {\mathbf e}_3\;{\mathbf e}_3
\right)
= \omega \,{\mathbf e}_3 + \lambda \, {\mathbf s}_0
\;.
\end{equation}
Note that the term $\lambda \, {\mathbf s}_0$ would have no influence on the eom
because of the vector product with ${\mathbf s}_0$.\\
The second term in (\ref{DC3}) already has the desired form:
\begin{eqnarray}\label{DC4a}
\sum_{n\in{\mathbf Z},\,n\neq 0}\frac{1}{|n|^3}
\left(-3\,{\mathbf s}_n\cdot {\mathbf e}_3\;{\mathbf e}_3
\right)
&=&
\sum_{n\in{\mathbf Z},\,n\neq 0}\frac{1}{|n|^3}
\left(-3\,z\;{\mathbf e}_3
\right)\\
\label{DC4b}
&=& -6 z \sum_{n=1}^\infty \frac{{\mathbf e}_3}{n^3} = -6\, z \,\zeta(3)\,{\mathbf e}_3
\;,
\end{eqnarray}
where $\zeta(n)$ denotes the Riemann Zeta function, see \cite{AS72}, 23.2.18.
To evaluate the first term in (\ref{DC3}) we add ${\mathbf s}_n$ and ${\mathbf s}_{-n}$ and obtain
\begin{eqnarray}\label{DC5a}
{\mathbf s}_n+{\mathbf s}_{-n}
&=& 2
\left(
\begin{array}{c}
 \sqrt{1-z^2} \cos (n \,q) \\
0 \\
 z \\
\end{array}
\right)\\
\label{DC5b}
&=&
2\, \cos (n \,q)
\left(
\begin{array}{c}
 \sqrt{1-z^2}  \\
0 \\
 z \\
\end{array}
\right)
+
\left(
\begin{array}{c}
0\\
0 \\
2 z(1-\cos(n \,q)) \\
\end{array}
\right)\\
\label{DC5c}
&=&
2\, \cos (n \,q)\,{\mathbf s}_0+2 z(1-\cos(n \,q))\,{\mathbf e}_3
\;.
\end{eqnarray}
Upon multiplying this result by $\frac{1}{n^3}$,summing over $n=1,\ldots,\infty$ and using (\ref{DC4b})
we obtain
\begin{eqnarray}\label{DC6a}
\omega&=& -6 z \zeta(3)+\sum_{n=1}^\infty \frac{1}{n^3}2 z(1-\cos(n \,q))\\
\label{DC6b}
&=&-z\left( 4 \zeta(3) +
\mbox{Li}_3\left(e^{-{\sf i} q}\right)+\mbox{Li}_3\left(e^{{\sf i} q}\right)\right)
\;.
\end{eqnarray}
Here $\mbox{Li}_n(z)$ denotes the polylogarithmic function, see \cite{PolyLog}. Eq.~(\ref{DC6b})
represents the dispersion relation $\omega=\omega(q,z)$ that also linearly depends on the parameter $z$.
For the comparison with the dispersion relations for infinitesimal spin waves and finite $N$
we have to set $z=1$ according to the ground state value, see Fig.~\ref{fig_dri}.\\
We note that for $z>0$ the dispersion relation (\ref{DC6b}) leads to a negative group velocity $\frac{d\omega}{dq}$.

\section{Summary and outlook}\label{sec:S}
In this article we have theoretically and numerically investigated spin waves in rings of fixed  dipoles
interacting solely via their magnetic fields. We have found numerical evidence for exact spin waves with finite amplitudes
and analytically solved three limit cases:
\begin{itemize}
  \item Spin waves that are infinitesimal excitations of the ground state,
  \item synchronized oscillations that can be viewed as spin waves with vanishing wave number and finite amplitudes, and
  \item spin waves with finite amplitudes in the infinite dipole chain.
\end{itemize}
The results for these three cases have been checked in various ways, e.~g.~, by numerical solutions of the eom including
Monte Carlo simulations for low temperatures, and by comparing the normalized dispersion relations of large rings with the corresponding
limit case of the infinite dipole chain. Thus we have completed the program arranged in the outlook of \cite{SSHL15} except
for the theoretical calculations of the low temperature asymptotics. Further topics that have to be deferred to forthcoming
papers are
\begin{itemize}
  \item the detailed numerical investigation of finite amplitude spin waves for small $N$, and
  \item a rigorous existence proof for finite amplitude spin waves in the general case.
\end{itemize}
The latter problem seems to be the most ambitious one and probably one will have to be content with numerical evidence.

Another question is how the theoretical results could be tested.
Note that our calculations are general and could be applied to microscopic as well as macroscopic realizations of interacting dipole systems. However, one has to make sure that higher order interactions, e.~g.~octopole, can be neglected. Furthermore, for microscopic systems it is important that the dipolar interaction is large due to large localized magnetic moments. This may be the case for special types of ring-type molecular magnets that are composed of rare earth metals with high spin quantum numbers like gadolinium \cite{UL12}. Another realization could be obtained by creating rings of interacting magnetic nanoparticles or lithographically produced micro- or nanostructures.

\section*{Acknowledgment}
We are indebted to Eva H\"agele and J\"urgen Schnack for valuable discussions about the subject of this article.

\appendix
\section{Ground states of the dipole ring}\label{sec:A}
We will introduce some notations and arguments referring to the ground states of the dipole ring without
formulating a rigorous proof. Such a proof can be found in \cite{S16}.
Obviously, the Hamiltonian (\ref{R4}) is bilinear in the components $s_{\mu i}$ of the moment vectors
${\mathbf s}_\mu$
and hence can be written in the form
\begin{eqnarray}\label{A1}
H&=&\sum_{\mu=0}^{N-1}\sum_{\nu=0}^{N-1}\sum_{i,j=1}^3 J_{\mu\nu i j} s_{\mu i}s_{\nu j}\\
&\equiv& \sum_{\alpha,\beta} {\mathbf J}_{\alpha\beta} s_\alpha s_\beta
\;,
\end{eqnarray}
where we have introduced multi-indices $\alpha=(\mu,i)\;,\beta=(\nu,j)$ that run through a finite set of size $3N$.
Let $j_{\mbox{\scriptsize min}}$ be the lowest eigenvalue of the symmetric matrix ${\mathbf J}$. Then,
by the Rayleigh-Ritz variation principle,
$H\ge \sum_{\alpha} j_{\mbox{\scriptsize min}} s_\alpha^2=N j_{\mbox{\scriptsize min}}$,
but the minimal energy $E_0$ need not be equal to $N j_{\mbox{\scriptsize min}}$ in general. In our case
there exists a certain eigenvalue $j_\alpha$ of ${\mathbf J}$ such that the corresponding eigenvector
can be identified with the state $\breve{\mathbf s}$, see (\ref{R5}). Hence $\breve{\mathbf s}$ is a ground state
if $j_\alpha=j_{min}$ since then the lower energy bound $N\, j_{min}$ would be assumed by the spin configuration $\breve{\mathbf s}$.
In this way the ground state problem is reduced to a matrix problem in close analogy to the Luttinger-Tisza approach \cite{LT46}.\\
To detail the above remarks it is convenient to introduce new cartesian coordinates
$(\xi,\eta,\zeta)\equiv(\xi_0,\ldots,\xi_{N-1},\eta_0,\ldots,\eta_{N-1},\zeta_0,\ldots,\zeta_{N-1})$
for the moment vectors ${\mathbf s}_\mu$ defined by
\begin{equation}\label{A2}
\xi_\mu={\mathbf r}_\mu\cdot{\mathbf s}_\mu,\;\eta_\mu={\breve{\mathbf s}}_\mu\cdot{\mathbf s}_\mu,\;,
\zeta_\mu={\mathbf e}_3\cdot{\mathbf s}_\mu,\quad \mu=0,\ldots,N-1\,.
\end{equation}
W.~r.~t.~these new coordinates that
are better adapted to the $C_N$-symmetry of the dipole ring
the matrix ${\mathbf J}$ assumes the form, again denoted by ${\mathbf J}$,
\begin{equation}\label{A3}
{\mathbf J}=
\left(
\begin{array}{ccc}
 A&C&0 \\
 -C&B&0 \\
 0&0&D \\
\end{array}
\right)
\;.
\end{equation}
Here $A,B,C,D$ denote $N\times N$ sub-matrices that are circulants, see section \ref{sec:L}.
$A,B$ and $D$ are symmetric, whereas $C$ is anti-symmetric. Again, the matrices $A,B,C,D$
pairwise commute since they have the Fourier basis ${\mathbf b}^{(\mu)}$, see (\ref{L3a}), as a common system of eigenvectors.
It thus suffices to give the entries of the first row of the respective matrices. For a more detailed derivation of these entries see \cite{S16}.
\begin{eqnarray}\label{A4a}
A_{0,\mu}&=&
\left\{
\begin{array}{l@{\;:\;}l}
 0& \mu=0\,,\\
\frac{1}{32} \left(3-\cos \left(\frac{2 \pi  \mu }{N}\right)\right) \csc ^3\left(\frac{\pi  \mu
   }{N}\right)
 & \mu=1,\ldots,N-1,
\end{array}
\right.\\
\label{A4b}
B_{0,\mu}&=&
\left\{
\begin{array}{l@{\;:\;}l}
 0& \mu=0\,,\\
-\frac{1}{32} \left(3+\cos \left(\frac{2 \pi  \mu }{N}\right)\right) \csc ^3\left(\frac{\pi  \mu
   }{N}\right)
 & \mu=1,\ldots,N-1,
\end{array}
\right.\\
\label{A4c}
C_{0,\mu}&=&
\left\{
\begin{array}{l@{\;:\;}l}
 0& \mu=0\,,\\
\frac{1}{32} \sin \left(\frac{2 \pi  \mu }{N}\right) \csc ^3\left(\frac{\pi  \mu }{N}\right)
 & \mu=1,\ldots,N-1,
\end{array}
\right.
\\
\label{A4d}
D_{0,\mu}&=&
\left\{
\begin{array}{l@{\;:\;}l}
 0& \mu=0\,,\\
\frac{1}{16} \csc ^3\left(\frac{\pi  \mu }{N}\right)
 & \mu=1,\ldots,N-1,
\end{array}
\right.
\end{eqnarray}

We see that, except a vanishing diagonal, $A$ and $D$  have only positive entries and $B$ has only negative ones.
The row sum of $C$ vanishes since $C$ is an anti-symmetric circulant.
The eigenvalues of $C$ are purely imaginary since it is also an anti-Hermitean matrix.
Let
\begin{equation}\label{Jnu}
  J^{(\mu)}\equiv\left(
\begin{array}{ccc}
 a^{(\mu)}& {\sf i}\,c^{(\mu)}&0 \\
 -{\sf i}\,c^{(\mu)}&b^{(\mu)}&0 \\
 0&0&d^{(\mu)} \\
\end{array}
\right)
\end{equation}
be defined as the $3\times 3$ matrix where $a^{(\mu)},b^{(\mu)}, {\sf i}\,c^{(\mu)},d^{(\mu)} $ are the eigenvalues of the corresponding sub-matrices
$A,B,C,D$ of ${\mathbf J}$ and $\mu=0,\ldots,N-1$.
Further let $j^{(\mu)}_i,\;i=1,2,3$ be the eigenvalues of $J^{(\mu)}$
with eigenvectors $u^{(\mu)}_i$.
Then the general eigenvector of $\mathbf J$ has the form
$(u^{(\mu)}_{i,1}{\mathbf b}^{(\mu)},u^{(\mu)}_{i,2}{\mathbf b}^{(\mu)},u^{(\mu)}_{i,3}{\mathbf b}^{(\mu)})^\top$
corresponding to the eigenvalue  $j^{(\mu)}_i$.

 In this way we have, in principle, diagonalized the matrix ${\mathbf J}$. In particular, its eigenvalues corresponding to $\mu=0$
 can be determined explicitely. The Fourier basis vector ${\mathbf b}^{(0)}$ is the vector with constant entries $\frac{1}{\sqrt{N}}$.
 The eigenvalues $a^{(0)}, b^{(0)}, {\sf i}\,c^{(0)}, d^{(0)}$ considered above are the constant row sums of $A,B,C,D$, where $c^{(0)}=0$, since $C$ has
 vanishing row sums. It follows that $J^{(0)}=\mbox{diag }(a^{(0)}, b^{(0)},  d^{(0)})$. Obviously, $b^{(0)}$ is the lowest eigenvalue of $J^{(0)}$
 since $b^{(0)}<0$ but $a^{(0)}>0$ and $d^{(0)}>0$. The corresponding eigenvector of ${\mathbf J}$ is
 $({\mathbf 0},{\mathbf b}^{(0)},{\mathbf 0})^\top$.
 It is, up to normalization, identical with the conjectured ground state $\breve{\mathbf s}$ according to (\ref{R5}). To prove that
 $\breve{\mathbf s}$ is actually a ground state it would suffice to show that $b^{(0)}$ is the lowest eigenvalue of ${\mathbf J}$
 since then the equality sign in $E_0\ge N j_{min}$ would be assumed.  For any given small $N$ this could be done since the eigenvalues
 of ${\mathbf J}$ can be calculated in closed form. A rigorous proof of the ground state property of $\breve{\mathbf s}$ following
 the above strategy would, however, require the proof of $b^{(0)}=j_ {min}$ for {\textit all} $N\ge 3$.\\

In the remainder of this Appendix we will present arguments in favour the ground state property of $\breve{\mathbf s}$ if $N$ is sufficient large.
These arguments can be developed further to yield a rigorous proof, see \cite{S16}.

First we will prove that
\begin{equation}\label{Perron1}
b^{(0)}<d^{(\nu)} \mbox{  for all } \nu=0,\ldots,N-1
\;.
\end{equation}
Recall that $d^{(\nu)}$ can be written as
\begin{eqnarray}\label{A5a}
d^{(\nu)}&=&\sum_{\mu=1}^{N-1}\frac{1}{16} \csc ^3\left(\frac{\pi  \mu }{N}\right) \exp\left(\frac{2\pi{\sf i}\nu\mu}{N} \right)\\
\label{A5b}
&=&\sum_{\mu=1}^{N-1}\frac{1}{16} \csc ^3\left(\frac{\pi  \mu }{N}\right) \cos\left(\frac{2\pi\nu\mu}{N} \right)
\;,
\end{eqnarray}
and hence
\begin{equation}\label{A6}
|d^{(\nu)}|\le \sum_{\mu=1}^{N-1}\frac{1}{16} \csc ^3\left(\frac{\pi  \mu }{N}\right)=d^{(0)}
\;,
\end{equation}
where the $=$ sign only applies for $\nu=0$.
Alternatively, we could have invoked the theorem of Perron (1907) in the form \cite{N76} in order to show (\ref{A6}).
Now we use the fact that $D_{0,\nu}<|B_{0,\nu}|$ for all $\nu=0,\ldots,N-1$ and hence $d^{(0)}<|b^{(0)}|$. Together with (\ref{A6})
this implies $|d^{(\nu)}|<|b^{(0)}|$ and further $b^{(0)}<d^{(\nu)}$ since $b^{(0)}<0$. \\
Similarly one also proves
\begin{equation}\label{Perron2}
b^{(0)}<b^{(\nu)} \mbox{  for all } \nu=0,\ldots,N-1
\;.
\end{equation}
Now we can restrict ourselves to the $2\times 2$ submatrix
$\left(
\begin{array}{cc}
 a^{(\nu)}& {\sf i}\,c^{(\nu)}\\
 -{\sf i}\,c^{(\nu)}&b^{(\nu)} \\
\end{array}
\right)$
of $J^{(\nu)}$.
We subtract $b^{(0)}$ in the diagonal and obtain the matrix
\begin{equation}\label{K}
K^{(\nu)}\equiv
\left(
\begin{array}{cc}
 a^{(\nu)}-b^{(0)}& {\sf i}\,c^{(\nu)}\\
 -{\sf i}\,c^{(\nu)}&b^{(\nu)}-b^{(0)} \\
\end{array}
\right)
\;.
\end{equation}
The ground state property of $\breve{\mathbf s}$ now follows if  $K^{(\nu)}$ is positive-definite,
i.~e.~if both eigenvalues of $K^{(\nu)}$ are strictly positive for all $\nu=0,\ldots,N-1$.
By virtue of Sylvester's criterion (positivity of all principal minors) and the positivity of $K^{(\nu)}_{22}$,
see (\ref{Perron2}), it remains to show that
\begin{equation}\label{det}
 \det K^{(\nu)}=(a^{(\nu)}-b^{(0)})(b^{(\nu)}-b^{(0)})-(c^{(\nu)})^2>0
 \;.
\end{equation}
After some elementary transformations we write $ \det K^{(\nu)}$ as a double sum of the form
\begin{eqnarray}
 \nonumber 
  \det K^{(\nu)} &=& \sum_{\lambda,\mu=1}^{N-1}k^{(\nu)}_{\lambda\mu}\\
  \nonumber
   &\equiv& \sum_{\lambda,\mu=1}^{N-1}
   \csc ^3\frac{\pi  \lambda }{N} \csc ^3\frac{ \pi  \mu }{N}\\
  \nonumber
  && \left[\left(1-\cos\frac{2 \pi  \mu  \nu   }{N}\right)
   \left(\cos \frac{2 \pi  \lambda   }{N}    \left(1-\cos \frac{2 \pi  \lambda  \nu   }{N}\right)\right.\right.\\
    \nonumber
    &&\left.\left.+3 \left(\cos\frac{2 \pi  \lambda  \nu}{N}  +1\right)\right)\right.\\
  \label{Adouble}
   &&\left.-\sin \frac{2 \pi  \lambda   }{N}   \sin \frac{2 \pi  \mu }{N}   \sin \frac{2 \pi  \lambda  \nu} {N}
    \sin  \frac{2 \pi  \mu  \nu }{N}\right]
    \;,
\end{eqnarray}
ignoring the irrelevant global factor $\left(\frac{1}{32}\right)^2$.
This can be viewed as a sum over a square lattice. If $N\longrightarrow\infty$ the terms at the boundary of the square lattice diverge
since, e.~g.~, $ \csc ^3\left(\frac{2 \pi  \lambda }{N}\right)=O(N^3)$ if $\lambda$ is small or close to $N$. The terms in the interior
of the square lattice remain bounded and at most contribute to $ k^{(\nu)}_{\lambda\mu}$ with an amount of order $O(N^2)$. In order to investigate the
leading terms of $ k^{(\nu)}_{\lambda\mu}$ in more detail we will consider the two cases $\nu=O(1)$ and $\nu=O(N)$.\\
Let us begin with the case $\nu=O(N)$ and consider only contributions from small $\lambda$ or $\mu$. The contributions from $\lambda$ or $\mu$
close to $N$ are completely analogous and would only give an irrelevant global factor of $4$. It turns out that the leading term is
due to the contribution of $\lambda=0(1)$ and $\mu=0(1)$ and reads:
\begin{equation}\label{A7}
  k^{(\nu)}_{\lambda\mu}=
  \frac{  \left(\cos\frac{2 \pi  \lambda  \nu}{N}+2\right) \sin ^2\frac{\pi  \mu  \nu }{N}}{\sqrt{2}\,\pi^6\, \lambda ^3\, \mu ^3}\,N^6+O(N^4)
   \;.
\end{equation}
Note that according to our assumption $\nu=O(N)$ the $\cos-$ and $\sin-$terms must not be expanded. Obviously, the leading term of (\ref{A7})
is positive. There are other contributions from the boundary of the square lattice, e.~g.~, if $\lambda=O(N)$  but $\mu=O(1)$. In this case
\begin{eqnarray}\nonumber
  k^{(\nu)}_{\lambda\mu}&=&
 \frac{ \csc ^3\frac{\pi  \lambda }{N}\left(1-\cos\frac{2 \pi  \mu  \nu }{N}\right) \left(\left(\cos\frac{2 \pi  \lambda }{N}
 -3\right) \sin ^2\frac{\pi\lambda\nu }{N}+3\right)}{2 \sqrt{2} \pi ^3 \mu ^3}N^3+O(N^2)
   \;.\\
  \label{A8}
\end{eqnarray}
This term may be negative, depending on $\lambda$, but even if it is multiplied by the maximal number of terms, $(N-1)^2$,
it will be dominated by the positive term (\ref{A7}) of order $O(N^6)$.
The leading contribution of $\lambda=O(1)$  but $\mu=O(N)$ is also of order $O(N^3)$ but always positive.
We conclude that $\det K^{(\nu)}>0$ if $N$ is sufficient large and $\nu=O(N)$.

Turning to the case $\nu=0(1)$ we find the following positive leading contribution due to $\lambda=0(1)$ and $\mu=0(1)$:
\begin{equation}\label{A9}
  k^{(\nu)}_{\lambda\mu}=
 \frac{48 \nu ^2}{\pi ^4 \lambda ^3 \mu }\,N^4+O(N^2)
   \;.
\end{equation}
The contributions from $\lambda=O(N)$  but $\mu=0(1)$ or $\lambda=0(1))$  but $\mu=O(N)$ are of order $O(N)$ and always positive. Hence
also in this case $\det K^{(\nu)}>0$ if $N$ is sufficiently large.


\section*{References}

\begin{thebibliography}{99}


\bibitem{EW13} Ewerlin M, Demirbas D, Br\"ussing F, Petracic O, \"Unal A A,
Valencia S, Kronast F and Zabel H 2013
\textit{Phys. Rev. Lett.} \textbf{110} 177209

\bibitem{MI13} Miloshevich G, Dauxois T, Khomeriki R and Ruffo S 2013
\textit{Eur. Phys. Lett.} \textbf{104} 17011

\bibitem{VA13} Varon M, Beleggia M, Kasama T, Harrison R J, Dunin-Borkowski R E, Puntes V F and Frandsen C
2013 \textit{Sci. Rep.} \textbf{3} 1234

\bibitem{DZ13a} Dzyan S A and Ivanov B A 2013 \textit{Low Temp. Phys.} \textbf{39} 525--529

\bibitem{DZ13b} Dzyan S A and Ivanov B A 2013
\textit{JETP} \textbf{116} 975--979

\bibitem{WU11} Wunsch B, Zinner N T, Mekhov I B, Huang S--J, Wang D--J and Demler E 2011
\textit{Phys. Rev. Lett.} \textbf{107} 073201

\bibitem{ST05} Stuhler J, Griesmaier A, Koch T, Fattori M, Pfau T, Giovanazzi S, Pedri P and Santos L
2005 \textit{Phys. Rev. Lett.} \textbf{95}  150406

\bibitem{UN05} Unold T, Mueller K, Lienau Ch, Elsaesser T and Wieck A D 2005
\textit{Phys. Rev. Lett.} \textbf{94} 137404

\bibitem{JO98} Jonsson T, Nordblad P and Svedlindh P 1998
\textit{Phys. Rev. B} \textbf{57} 497--594


\bibitem{PS16} Pohlit M, Stockem I, Porrati F, Huth M,  Schr\"oder C,  M\"uller J
2016 \textit{J. Appl. Phys.} \textbf{120} 142103

\bibitem{VdW16} Van de Wiele B, Fin S, Pancaldi M, Vavassori P, Sarella A, Bisero D 2016 \textit{J. App. Phys.} \textbf{119} 203901

\bibitem{MCT16} Martinez-Pedrero F, Cebers A, and Tierno P
2016
\textit{Phys. Rev. Applied} \textbf{6} 034002




\bibitem{W06} Wang R F, Nisoli C, Freitas R S, Li J, McConville W,
Cooley B J, Lund M S, Samarth N, Leighton C, Crespi V H and P.~Schiffer P
2006 \textit{Nature} \textbf{439}  303-306

\bibitem{C08}Castelnovo C, Moessner R and Sondhi S L 2008 \textit{Nature} \textbf{451} 06433


\bibitem{H11} Hiroi K, Komatsu K and Sato T 2011 \textit{Phys. Rev. B} \textbf{83} 224423

\bibitem{I06} Imre A, Csaba G, Ji L, Orlov A, Bernstein G H and Porod W 2006 \textit{Science} \textbf{311} 205-208

\bibitem{E14} Eichwald I, Breitkreutz S, Ziemys G, Csaba G, Porod W and Becherer M 2014
\textit{Nanotechnology} \textbf{25} 335202

\bibitem{W00} Waiblinger M, Goedde B, Jakes P, Lips K, Harneit W, Weidinger A and Dinse K P 2000
\textit{AIP Conference Proceedings} \textbf{544}  203-206

\bibitem{H02} Harneit W 2002 \textit{Phys. Rev. A}  \textbf{65} 032322


\bibitem{SSHL15} Schmidt H-J, Schr\"oder C, H\"agele E and Luban M
2015
\textit{J. Phys. A} \textbf{48} 185002

\bibitem{SSR15} Sch\"onke J, Schneider M and, Rehberg I
2015
\textit{Phys. Rev. B} \textbf{91} 020410


\bibitem{S16} Schmidt H-J 2016
\verb" arXiv: 1610.09309"



\bibitem{A97} Anderson P W 1997
{\it Basic notions of condensed matter physics} (Addison-Wesley: Reading, Mass)

\bibitem{M81} Mattis D C 1981
{\it Theory of magnetism I} (Springer: Berlin)


\bibitem{SSL11} Schmidt H-J, Schr\"oder C and Luban M
2011
\textit{J. Phys.: Condens. Matter} \textbf{23} 386003


\bibitem{AT16} Amano H and Tomita M
2016
\textit{Phys. Rev. A} \textbf{93} 063854

\bibitem{LX16} Li X-J, Xue C, Fan L, et al.
2016
\textit{Appl. Phys. Lett.} \textbf{108} 231904

\bibitem{MG16}  Mruczkiewicz M, Gruszecki P,  Zelent M, et al.
2016
\textit{Phys. Rev. B} \textbf{93} 174429


\bibitem{G99} Griffith D J 1999
{\textit Introduction to Electrodynamics} 3rd ed. (Upper Saddle River, New Jersey: Prentice Hall)


\bibitem{VD14} Vandewalle N and Dorbolo S 2014 \textit{New Journal of Physics} \textbf{16} 013050

\bibitem{LL99} Luban M and Luscombe J H 1999 \textit{Am. J. Phys.} \textbf{67} 1161--1169

\bibitem{A01} Aldrovandi R 2001
{\it Special Matrices of Mathematical Physics} (World Scientific: Singapore)


\bibitem{A78} Arnol'd V I 1978
{\it Mathematical Methods of Classical Mechanics} (Springer: Berlin)

\bibitem{AS72} Abramowitz M and Stegun I A 1972 (eds)
{\it Handbook of Mathematical Functions} (New York: Dover)

\bibitem{PolyLog}
\verb" http://mathworld.wolfram.com/Polylogarithm.html",

\bibitem{UL12}  Ungur L, Langley S, Hooper T, Moubaraki B, Brechin E K, Murray K S, and Chibotaru L 2012
\textit{ J. Am. Chem. Soc} \textbf{6134}, no. 45, 18554 - 18557

\bibitem{LT46} Luttinger JM and Tisza L 1946
\textit{Phys. Rev.} \textbf{70} 954 -- 964

\bibitem{N76}
F. Ninio,  J. Phys. A {\bf 9}  No.  8, 1281 (1976)





\end{thebibliography}
\end{document}